\documentclass[11pt,reqno]{amsart}
\numberwithin{equation}{section}
\usepackage{amsmath}
\usepackage{amssymb}
\usepackage{amsfonts}
\usepackage[round]{natbib} 
\usepackage{xcolor}
\usepackage{graphicx}
\usepackage{subfigure}
\usepackage{stmaryrd}
\usepackage{caption}
\usepackage{import}
\usepackage{xifthen}
\usepackage{pdfpages}
\usepackage{transparent}

\usepackage{comment}
\usepackage{tikz}
\usepackage{url}
\usepackage[linktocpage]{hyperref}
\hypersetup{
    colorlinks=true,
    citecolor=black,
    linkcolor=black,
    filecolor=black,
    urlcolor=black,}
\usepackage{todonotes}
\calclayout
\begin{document}

\title[2-D turbulence in ideal magnetized fluids]{Structure-preserving long-time simulations of turbulence in magnetized ideal fluids}
\author{Klas Modin}
\address{Klas Modin: Department of Mathematical Sciences, Chalmers University of Technology and University of Gothenburg, 412 96 Gothenburg, Sweden}
\email{klas.modin@chalmers.se}

\author{Michael Roop}
\address{Michael Roop: Department of Mathematical Sciences, Chalmers University of Technology and University of Gothenburg, 412 96 Gothenburg, Sweden}
\email{michael.roop@chalmers.se}



\begin{abstract}
We address three two-dimensional magnetohydrodynamics models: reduced magnetohydrodynamics (RMHD), Hazeltine's model, and the Charney--Hasegawa--Mima (CHM) equation.
    These models are derived to capture the basic features of magnetohydrodynamic turbulence and plasma behaviour.
    They all possess non-canonical Hamiltonian formulations in terms of Lie--Poisson brackets, which imply an infinite number of conservation laws along with symplecticity of the phase flow.
    This geometric structure in phase space affects the statistical long-time behaviour.
    Therefore, to capture the qualitative features in long-time numerical simulations, it is critical to use a discretization that preserves the rich phase space geometry.
    Here, we use the matrix hydrodynamics approach to achieve structure-preserving discretizations for each model.
    We furthermore carry out long-time simulations with randomized initial data and a comparison between the models.
    The study shows consistent behaviour for the magnetic potential: both RMHD and Hazeltine's model produce magnetic dipoles (in CHM, the magnetic potential is prescribed).
    These results suggest an inverse cascade of magnetic energy and of the mean-square magnetic potential, which is empirically verified via spectral scaling diagrams.
    On the other hand, the vorticity field dynamics differs between the models:
    RMHD forms sharp vortex filaments with rapidly growing vorticity values, whereas Hazeltine's model and CHM show only small variation in the vorticity values.
    Related to this observation, both Hazeltine's model and CHM give spectral scaling diagrams indicating an inverse cascade of kinetic energy not present in RMHD.
\end{abstract}

\maketitle

\section{Introduction}
% \printinunitsof{cm}\prntlen{\textwidth} \prntlen{\textheight}
Reduced magnetohydrodynamics (RMHD) serves as a simple model for both astrophysical and laboratory plasmas.
It models the motion of an incompressible conductive fluid consisting of charged particles in an ambient magnetic field. 
It was developed by \citet{Strauss1976} for low-beta dynamics (dominance of the magnetic pressure $p_{m}=B^{2}$ over the kinetic pressure $p=nk_{B}T$) of plasma in tokamaks. 
In this reduced setting, the plasma is assumed to be nearly two-dimensional, inviscid, and incompressible.

\begin{figure}
    \centering
    \includegraphics[scale=0.21]{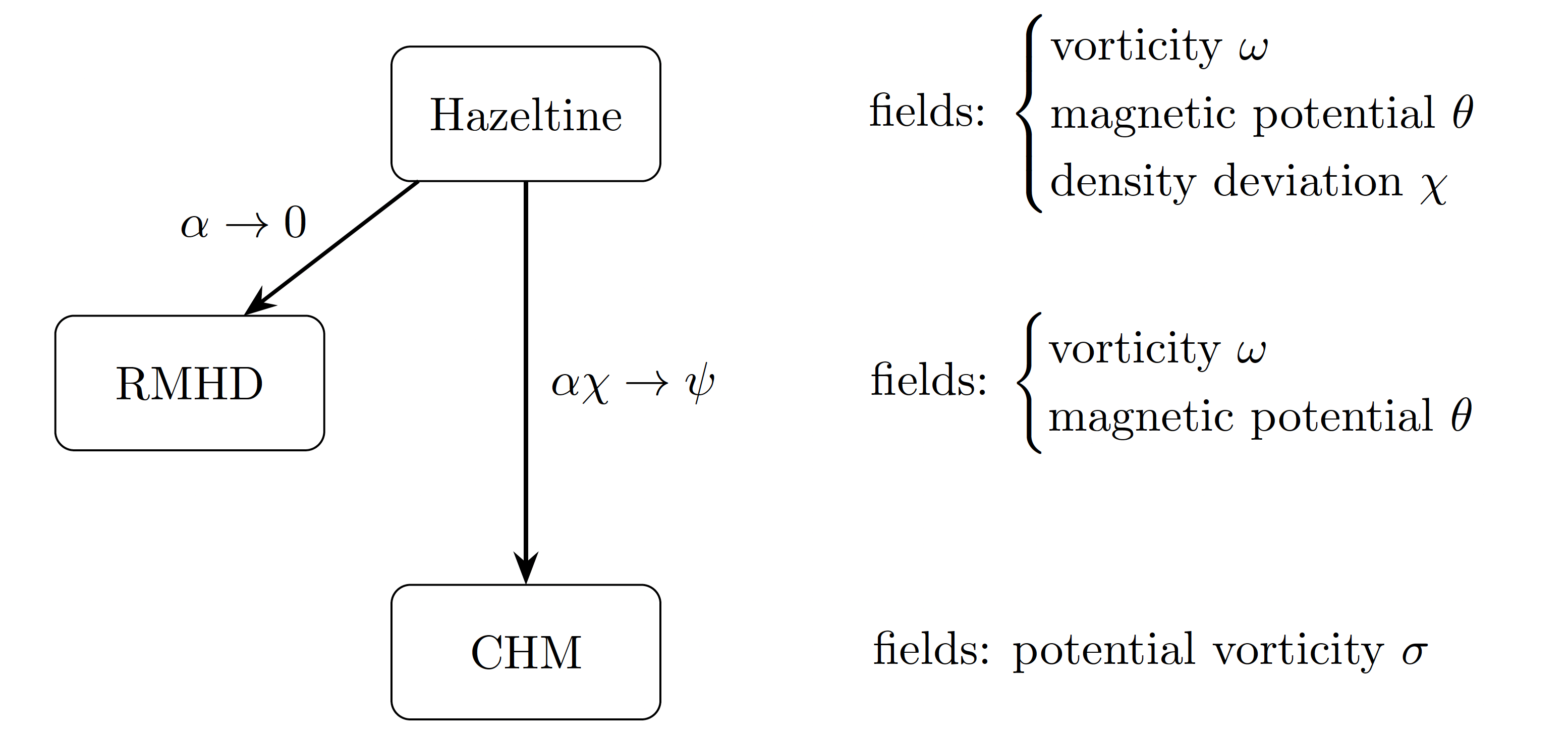}
    \caption{Overview of the relation between reduced models of magnetohydrodynamics.}
    \label{fig:overview}
\end{figure}

The RMHD model was generalized by \citet{Haz} to include the density variation, resulting in a three-field model.
Furthermore, \citet{CHM1977} introduced a model to describe electrostatic plasma turbulence.
The same model was also considered by \citet{Charney1948} in the context of geophysical fluid dynamics.
The relation between the two-field RMHD model, the three-field Hazeltine model, and the single-field Charney--Hasegawa--Mima (CHM) model is the following:
Hazeltine's model contains both RMHD and CHM as limiting cases (see Figure~\ref{fig:overview}). Further generalizations of RMHD include, for example, electron inertia effects \citep{SPK1994}, a four-field model for tokamak dynamics \citep{HHM1987}, and an inclusive model for Hall and inertia MHD \citep{GTAM2017}. In this paper, however, we restrict our attention to the three aforementioned models in Figure~\ref{fig:overview}.

The RMHD system and its extensions admit non-canonical Hamiltonian formulations in terms of Lie--Poisson structures on duals of semidirect product Lie algebras~\citep{MoGr1980,HazMorr,HolmKuper1983}. 
Such Hamiltonian formulations reveal the existence of conserved quantities, such as energy and Casimir invariants, along with symplecticity of the phase flow.
% called \textit{Casimirs}. 
The RMHD system has two families of Casimir invariants, parameterized by two arbitrary smooth functions and referred to as \textit{magnetic helicity} and \textit{cross-helicity}. 
The Hazeltine model, whose non-canonical Hamiltonian formulation was explored by \citet{HazHolm,Holm}, inherits the magnetic helicity invariant from RMHD, but not the cross-helicity invariant.
Instead, the system has another family of Casimir invariants parameterized by two arbitrary smooth functions (see the details below).
The discovery of the Lie--Poisson structure of magnetohydrodynamic models has made it possible to extend Arnold's stability analysis for ideal fluids \citep{ArnoldVI1969} to plasma physics models and to find new plasma equilibria \citep{HazHolm}, as well as sufficient conditions of their stability \citep{HMRW}.

The Lie--Poisson structure and its associated Hamiltonian dynamics impose geometrical properties in phase space that affect the long-time statistical behaviour (an example is the inverse energy cascade in 2-D hydrodynamics).
Thus, in the discretization of Lie--Poisson systems it is natural to aim for numerical schemes that preserve as much as possible of the structure, to retain the right statistical behaviour. In particular, the Lie--Poisson structure imposes invariant submanifolds in phase space called \textit{co-adjoint orbits}, and the dynamics is confined to these orbits (for details on Lie--Poisson dynamics we refer to the monograph by~\citet{MarsRat}). 
However, traditional numerical schemes, such as finite element, finite volume, or spectral methods, are typically designed to provide high accuracy in terms of minimizing the local error, but they fail to preserve the geometric structure, particularly the confinement to co-adjoint orbits.
For long-time simulations of chaotic systems, such as those discussed in this paper, high accuracy is less important, because small perturbations grow exponentially (or faster).
Instead, to retain the statistically correct qualitative behaviour, it is the phase space structure (such as co-adjoint orbits) that is important to capture numerically. Indeed, it is the geometrical structure, not the local dynamics, that is responsible for the formation of large scale structures in the long-time behaviour, such as the presence of the inverse cascade of energy in the case of hydrodynamical turbulence, and of the mean square of magnetic potential in MHD turbulence. 
One way to see this is via statistical mechanics, which is surprisingly useful for predicting the long-time behaviour, and which disregards the local dynamics altogether, but is directly affected by the geometric structure (in particular, the conservation laws and the symplectic nature of the phase flow). 
\textit{Geometric numerical methods}, i.e., numerical discretizations that preserve the underlying phase space geometry offer a middle ground: they take the geometry into account, while still approximating the local dynamics.\footnote{Although it is sometimes possible to construct geometric methods with as high order as traditional methods, it is usually more efficient to use low-order geometric methods, since the important property is conservation of phase space geometry, not high accuracy in tracking individual trajectories.}

For RMHD models, a structure preserving discretization was developed by \citet{KrTassGrass2016} based on a discrete variational principle. 
Their method preserves energy and the quadratic Casimir, but not the Lie--Poisson structure or the higher-order Casimirs.

Another geometric discretization algorithm was developed by the authors~\citep{ModRoop}. 
The key idea is to replace the infinite-dimensional Lie--Poisson system with a finite-dimensional matrix analogue, following the approach of \citet{Zeit,Ze1991,Ze2005} who, based on geometric quantization~\citep{Hopp,Hopp1,Bord1991,Bord1994,HoppYau}, developed matrix approximations for ideal fluids and RMHD on the flat torus (the doubly periodic square). 
The matrix dynamical system studied by~\citet{ModRoop} is itself a Lie--Poisson flow for a modified Lie--Poisson structure and retains (discrete versions of) the conservation laws, such as Casimir functions (magnetic helicity and cross-helicity) and energy. 
Furthermore, the modified Lie--Poisson structure, induced by the matrix equations, converge, along with the Casimirs, to the infinite-dimensional Lie--Poisson structure as the matrix dimension increases. 
As a full spatio-temporal discretization, the method gives a ``magnetic'' extension of the structure preserving matrix hydrodynamics approach to ideal fluids on the sphere \citep{ModViv,ModViv1}. 
Although the method is specifically focused on the case of the sphere, the matrix equations for spherical RMHD are modelled on the same Lie algebra as those on the flat torus, so the time integrator can also be applied to RMHD on the torus. 
The sphere is, however, the preferred choice for simulation of Zeitlin's equations because it fully preserves the invariance under the group of isometries, contrary to the corresponding matrix model for the flat torus (see the paper by \citet{ModinVivianiConvEu} for details). 

The aim of the present paper is (i) to numerically investigate basic properties of long-time turbulence of magnetic fluids on the sphere as modelled by the RMHD equations, Hazeltine's model, and the CHM model, and (ii) to compare the resulting long-time behaviour between the models. 
A motivation for our study comes from previous simulation results on hydrodynamic turbulence on the sphere \citep{CiViMo2023,ModViv1}, which demonstrate the formation of large-scale condensates. 
Do formations like those appear also in the various models for reduced magnetized fluids?

The paper is organized as follows. First, in Sect.~\ref{sect2}, we discuss the Hamiltonian formulation of incompressible MHD in terms of non-canonical Poisson brackets and corresponding Casimirs invariants. Further, in Sect.~\ref{sect3}, we provide structure preserving spatio-temporal discretization for incompressible MHD on the sphere. In Sect.~\ref{sec:long_time_simulations}, we perform a numerical study of the long-time behaviour of the three models. We conclude the paper in Sect.~\ref{sect_concl}.

\medskip

\textbf{Acknowledgements.} 
The authors would like to thank Darryl D. Holm and Philip J. Morrison for pointing us to Hazeltine's model for magnetic fluids.

\medskip

\textbf{Funding.}
This work was supported by the Swedish Research Council (grant number 2022-03453); the Knut and Alice Wallenberg Foundation (grant number WAF2019.0201); the Göran Gustafsson Foundation for Research in Natural Sciences and Medicine; and the Royal Swedish Academy of Sciences (M.R., grant numbers MA2024-0034, MG2024-0050, MA2025-0073). 
The computations were enabled by resources provided by Chalmers e-Commons at Chalmers.
The data handling was enabled by resources provided by the National Academic Infrastructure for Supercomputing in Sweden (NAISS), partially funded by the Swedish Research Council (grant  number 2022-06725).

\section{Reduced models of magnetized fluids}
\label{sect2}
We begin with a description of the two-dimensional models of ideal magnetohydrodynamics, namely RMHD, Hazeltine's model, and CHM, in terms of their non-canonical Hamiltonian structures and associated Casimir invariants.

\subsection{Reduced magnetohydrodynamics (RMHD)}
Let $(M,g)$ be a compact, two-dimensional Riemannian manifold without boundary.
% , with the trivial first cohomology group. 
The RMHD model on $(M,g)$ reads
\begin{equation}
\label{MHDvort}
\left\{
\begin{aligned}
&\dot\omega=\left\{\omega,\psi\right\}+\left\{\theta,j\right\}, \quad &\omega =\Delta\psi, \\
&\dot\theta=\left\{\theta,\psi\right\}, \quad &j=\Delta\theta, 
\end{aligned}
\right.
\end{equation}
where $\omega$ is the vorticity, $\psi$ is the stream function, $\theta$ is the magnetic potential, $j$ is the current density, all depending on time $t$ and $(x,y)\in M$. The Laplace--Beltrami operator $\Delta$ is given in terms of the metric $g$, and the Poisson bracket $\left\{\cdot,\cdot\right\}$ is defined via the symplectic form (or area-form) $\Omega$ on $M$ as
\begin{equation*}
\left\{f,h\right\}=\Omega(X_{f},X_{h}),\quad f,h\in C_{0}^{\infty}(M),
\end{equation*}
where $X_{f}$, $X_{h}$ are Hamiltonian vector fields with Hamiltonian functions $f$ and $h$ respectively. 
Since $M$ is a compact manifold without a boundary, one can normalize all the fields $\omega,j,\theta,\psi$ to have zero-mean, i.e., they are elements of $C_{0}^{\infty}(M)$. 

The system \eqref{MHDvort} possesses a non-canonical Hamiltonian formulation in terms of a Lie--Poisson structure on the dual of a semidirect product Lie algebra. The Hamiltonian for \eqref{MHDvort} is
\begin{equation}
\label{hamRMHD}
H=\frac{1}{2}\int\limits_{M}\left(|\nabla\psi|^{2}+|\nabla\theta|^{2}
\right)\mathrm{d}x\mathrm{d}y = -\frac{1}{2}\int\limits_{M}(\omega\psi+\theta j)\mathrm{d}x\mathrm{d}y ,
\end{equation}
where the first term is the kinetic energy, and the second term is the magnetic energy. The Lie--Poisson bracket
\begin{equation}
\label{RMHDbracket}
\begin{split}
 \llbracket F,G\rrbracket&=\int\limits_{M}\left[\omega\left\{\frac{\delta F}{\delta \omega},\frac{\delta G}{\delta \omega}\right\}+
 \theta\left(\left\{\frac{\delta F}{\delta \theta},\frac{\delta G}{\delta \omega}\right\}+\left\{\frac{\delta F}{\delta \omega},\frac{\delta G}{\delta \theta}\right\}\right)\right]\mathrm{d}x\mathrm{d}y
 \end{split}
\end{equation}
is called the \textit{semidirect product bracket} \citep{HMR}. The RMHD system \eqref{MHDvort} then reads
\begin{equation*}
\dot F=\llbracket F,H\rrbracket,
\end{equation*}
where $F$ is an observable of the fields $\omega$ and $\theta$.

System \eqref{MHDvort} has an infinite number of invariants parameterized by two arbitrary smooth functions $f\colon\mathbb{R}\to\mathbb{R}$ and $g\colon\mathbb{R}\to\mathbb{R}$:
\begin{equation}
\label{cascont}
\mathcal{C}_{f}=\int\limits_{M}f(\theta)\mathrm{d}x\mathrm{d}y,\quad\mathcal{I}_{g}=\int\limits_{M}\omega g(\theta)\mathrm{d}x\mathrm{d}y.
\end{equation}
These invariants are Casimirs, i.e., for any functional $\mathcal{J}(\omega,\theta)$, one has $\llbracket \mathcal{C}_{f},\mathcal{J}\rrbracket=\llbracket \mathcal{I}_{g},\mathcal{J}\rrbracket=0$. 
We observe that $g(\theta)=\theta$ corresponds to the cross-helicity invariant, and $f(\theta)=\theta^{2}$ corresponds to the mean-square magnetic potential, both critical in MHD turbulence. 
More generally, we refer to $\mathcal{C}_{f}$ as magnetic helicity and to $\mathcal{I}_{g}$ as cross-helicity.
Of course, the Hamiltonian \eqref{hamRMHD} is also conserved by the flow \eqref{MHDvort}, but it is not a Casimir invariant. Instead, its conservation follows from the anti-symmetry of the Lie--Poisson bracket.

\subsection{Hazeltine's model and the Charney-Hasegawa-Mima (CHM) model}
A three-field extension of the RMHD system~\eqref{MHDvort} is given by the system of equations derived by \citet{Haz}
\begin{equation}
\label{Alfven}
\left\{
\begin{aligned}
&\dot\omega=\left\{\omega,\psi\right\}+\left\{\theta,j\right\}, \\
&\dot\theta=\left\{\theta,\psi\right\}-\alpha\left\{\theta,\chi\right\},\\
&\dot\chi=\left\{\chi,\psi\right\}+\left\{\theta,j\right\},
\end{aligned}
\right.
\end{equation}
where the new field $\chi$ represents the normalized deviation of the plasma density from a constant equilibrium value. 
The $\chi$ field is coupled to $\omega$ and $\theta$ via the length-scale parameter $\alpha\geq 0$.

Particular cases of equations \eqref{Alfven} include both low-beta limit of ideal RMHD and the CHM equation \citep{Charney1948,CHM1977}. 
First, the RMHD theory is recovered from \eqref{Alfven} by setting $\alpha=0$. 
Indeed, the system then decouples the $\omega$ and $\theta$ fields from the $\chi$ field, and the resulting equations constitute the RMHD system with the extra ``passive'' field $\chi$.
Second, the CHM equation is obtained from \eqref{Alfven} by presuming $\alpha\chi=\psi$. In this case, $\dot\theta=0$, and \eqref{Alfven} simplifies to the equation
\begin{equation}
\frac{\partial}{\partial t}\left(\omega-\frac{\psi}{\alpha}\right)+\left\{\psi,\omega\right\}=0,\quad\omega=\Delta\psi,
\end{equation}
where $\psi(t,x,y)$ is the velocity stream function. 
Introducing the new field, \textit{potential vorticity}, as $\sigma=\omega-\psi/\alpha$, we get the transport equation for the field $\sigma$ related to the stream function $\psi$ via the homogeneous Helmholtz operator:
\begin{equation}
\label{CHM}
\dot \sigma=\left\{\sigma,\psi\right\},\quad \sigma=\Delta\psi-\psi/\alpha
\end{equation}
The Casimir invariants for the system~\eqref{CHM} are
\begin{equation}
\label{CHMcas}
\mathcal{C}_{f}=\int\limits_{M}f(\sigma)\mathrm{d}x\mathrm{d}y,
\end{equation}
for arbitrary $f$ as before.
The equation \eqref{CHM} also arises in geophysical fluid dynamics \citep{LarRez1976}.

The Lie-Poisson bracket for Hazeltine's equations~\eqref{Alfven} is \citep{HazHolm,Holm}
\begin{equation}
\label{alfbracket}
\begin{split}
 \llbracket F,G\rrbracket&=\int\limits_{M}\left[\omega\left\{\frac{\delta F}{\delta\omega},\frac{\delta G}{\delta\omega}\right\}+
 \theta\left(\left\{\frac{\delta F}{\delta\theta},\frac{\delta G}{\delta\omega}\right\}+\left\{\frac{\delta F}{\delta\omega},\frac{\delta G}{\delta\theta}\right\}\right)+\right.{}\\&\left.
 \chi\left(\left\{\frac{\delta F}{\delta\chi},\frac{\delta G}{\delta\omega}\right\}+\left\{\frac{\delta F}{\delta\omega},\frac{\delta G}{\delta\chi}\right\}\right)+
 \chi\left\{\frac{\delta F}{\delta\chi},\frac{\delta G}{\delta\chi}\right\}+\right.{}\\&\left.
 \theta\left(\left\{\frac{\delta F}{\delta\theta},\frac{\delta G}{\delta\chi}\right\}+\left\{\frac{\delta F}{\delta\chi},\frac{\delta G}{\delta\theta}\right\}\right)\right]\mathrm{d}x\mathrm{d}y,
 \end{split}
\end{equation}
and the Hamiltonian function is
\begin{equation}
\label{hamcont}
H=-\frac{1}{2}\int\limits_{M}\left(\omega\psi+\theta j-\alpha\chi^{2}\right)\mathrm{d}x\mathrm{d}y.
\end{equation}
Thus, Hazeltine's equations can be written
\begin{equation*}
\dot F=\llbracket F,H\rrbracket,
\end{equation*}
for an observable $F$ of the fields $\omega$, $\theta$, and $\chi$.
Furthermore, the corresponding Casimir invariants are
\begin{equation*}
\mathcal{C}_{f}=\int\limits_{M}f(\theta)\mathrm{d}x\mathrm{d}y,\quad \mathcal{I}_{g}=\int\limits_{M}\chi g(\theta)\mathrm{d}x\mathrm{d}y,\quad\mathcal{P}_{k}=\int\limits_{M}k(\omega-\chi)\mathrm{d}x\mathrm{d}y,
\end{equation*}
for arbitrary smooth functions $f,g,k$.

We observe that all the three models have different Casimir functions. In particular, the cross-helicity Casimir $\mathcal{I}_{g}$ in \eqref{cascont} is not preserved by \eqref{Alfven}, and \eqref{CHMcas} is not preserved by either of \eqref{Alfven} and \eqref{MHDvort}.

\section{Matrix discretizations}
\label{sect3}
In this section, we give spatio-temporal discretizations for the systems \eqref{MHDvort} and \eqref{Alfven} based on finite-dimensional, matrix versions of the Lie--Poisson structures discussed previously. 
The main quality of such discretizations is that they capture all the geometric structure in the problem, rather than focusing on local accuracy as traditional numerical discretizations do. 
% To that end, the infinite-dimensional Lie--Poisson structures \eqref{RMHDbracket} and \eqref{alfbracket} are approximated by their matrix analogs, 
The idea goes back to \citet{Ze1991,Zeit,Ze2005} and is based on geometric quantization theory~\citep{Hopp,Hopp1,Bord1991,Bord1994,HoppYau}. 

From here on, we fix the manifold $M$ to be the two-sphere, $M=S^{2}$. 
Although Zeitlin's model was originally developed for the flat torus, it performs better on the sphere because of its consistency with the $SO(3)$ symmetry of the sphere (\emph{cf.}~\citet{ModinVivianiConvEu}). Albeit most MHD simulations in the literature are on the flat torus (doubly periodic square), we view the sphere as an equally suitable domain to investigate basic properties of MHD turbulence.

\subsection{The RMHD--Zeitlin equations}
The idea of finding a structure-preserving discretization is based on quantization theory, namely to replace the infinite-dimensional Poisson algebra of smooth functions $(C^{\infty}_{0}(S^{2}),\left\{\cdot,\cdot\right\})$ with a finite-dimensional matrix approximation consisting of the Lie algebra of skew-Hermitian matrices $\mathfrak{su}(N)$ equipped with the scaled Lie bracket $[\cdot,\cdot]_{N}=\frac{1}{\hbar}[\cdot,\cdot]$ for $\hbar=2/\sqrt{N^{2}-1}$.
As $N\to\infty$, the matrix Lie algebra $(\mathfrak{su}(N),[\cdot,\cdot]_N)$  converges, via the spherical harmonics basis (see below), to the Poisson algebra $(C^{\infty}_{0}(S^{2}),\left\{\cdot,\cdot\right\})$ of zero-mean smooth functions on $S^{2}$ (see, for example, the convergence result of \citet{ChPolt2017}).
A finite-dimensional approximation for the RMHD equations~\eqref{MHDvort} is therefore given by the following \emph{RMHD--Zeitlin equations}
\begin{equation}
\label{qMHDvort}
\left\{
\begin{aligned}
&\dot W=\frac{1}{\hbar}[W,P]+\frac{1}{\hbar}[\Theta,J],\quad &W=\Delta_{N}P \\
&\dot\Theta=\frac{1}{\hbar}[\Theta,P],\quad &J=\Delta_{N}\Theta,
\end{aligned}
\right.
\end{equation}
where $W,P,J,\Theta\in\mathfrak{su}(N)$ are matrices and $\Delta_{N}\colon\mathfrak{su}(N)\to\mathfrak{su}(N)$ is the \textit{Hoppe--Yau Laplacian} \citep{HoppYau}. 

To obtain the connection between matrices, such as $W$, and fields, such as $\omega$, we consider the eigen-matrices of the Hoppe--Yau Laplacian $\Delta_N$, which form a basis $T_{lm}^{N}$, $l=0,1,\ldots, N-1,\,m=-l,\ldots,l$ for $\mathfrak{su}(N)$ called \textit{matrix harmonics}. 
Then, for $W\in\mathfrak{su}(N)$, one reconstructs the vorticity function $\omega\in C^{\infty}_{0}(S^{2})$, up to the first $N^{2}$ terms of its Fourier decomposition in the spherical harmonics basis $Y_{lm}$, by the correspondences $T_{lm}^{N} \sim Y_{lm}$. 
For more details on mapping between the vorticity matrix $W\in\mathfrak{su}(N)$ and the vorticity function $\omega\in C^{\infty}_{0}(S^{2})$, we refer to the publications by~\citet{ModinVivianiConvEu,ModRoop}. 

The system \eqref{qMHDvort} constitutes a finite-dimensional Lie--Poisson flow on the dual $\mathfrak{f}^{*}$ of the semidirect product Lie algebra $\mathfrak{f}=\mathfrak{su}(N)\ltimes\mathfrak{su}(N)^{*}$. 
The integral of a function on $S^2$ corresponds to $4\pi/N$ times the trace of the corresponding matrix.
In particular, the Hamiltonian for \eqref{qMHDvort} is obtained from the Hamiltonian \eqref{hamRMHD} as
\begin{equation}
\label{qHamiltMHD}
H(W,\Theta)=\frac{2\pi}{N}\mathrm{tr}\left(W^{\dag}P + \Theta^{\dag}J \right).
\end{equation}
Likewise, the discretized Casimir invariants are
\begin{equation}
\label{qCas}
\mathcal{C}_{f}^{N}=\frac{4\pi}{N}\mathrm{tr}\left(f(\Theta)\right),\quad \mathcal{I}_{g}^{N}=\frac{4\pi}{N}\mathrm{tr}(Wg(\Theta)).
\end{equation}
As $N\to\infty$, the Hamiltonian \eqref{qHamiltMHD} and the Casimirs \eqref{qCas} converge to their continuous counterparts \eqref{hamRMHD} and \eqref{cascont} (for $f$ and $g$ taken as monomials, see the paper by \citet{ModRoop} for the exact result).

The matrix flow \eqref{qMHDvort} is integrated in time using the Lie--Poisson  preserving time integrator $\Phi_{h}\colon(W_{n},\Theta_{n})\mapsto(W_{n+1},\Theta_{n+1})$ defined by the equations %depending on the time step size $h$,
\begin{equation}
\label{MHDmethod1}
\begin{aligned}
&\Theta_{n}=\tilde\Theta-\frac{\varepsilon}{2}[\tilde\Theta,\tilde{P}]-\frac{\varepsilon^{2}}{4}\tilde{P}\tilde\Theta\tilde{P}, \\
&\Theta_{n+1}=\Theta_{n}+\varepsilon[\tilde\Theta,\tilde{P}],\\
&W_{n}=\tilde W-\frac{\varepsilon}{2}[\tilde W,\tilde{P}]-\frac{\varepsilon}{2}[\tilde\Theta,\tilde J]-\frac{\varepsilon^{2}}{4}\left(\tilde{P}\tilde{W}\tilde{P}+\tilde J\tilde\Theta\tilde{P}+\tilde{P}\tilde\Theta\tilde J\right), \\
&W_{n+1}=W_{n}+\varepsilon[\tilde W,\tilde P]+\varepsilon[\tilde\Theta,\tilde J],
\end{aligned}
\end{equation}
where $\tilde P=\Delta_{N}^{-1}\tilde W$, $\tilde J=\Delta_{N}\tilde\Theta$, and $\varepsilon = \delta t/\hbar$ for the physical time step length $\delta t > 0$.
The integrator \eqref{MHDmethod1} exactly\footnote{Up to the machine precision of floating-point numbers when implemented on a computer.} preserves the Casimirs~\eqref{qCas} for any choice of functions $f$ and $g$, and nearly preserves the Hamiltonian~\eqref{qHamiltMHD}, see \citet{ModRoop}. 

The scheme \eqref{MHDmethod1} is implicitly defined, requiring root-finding for the pair $(\tilde W,\tilde\Theta)$ as an intermediate step. This is achieved by fixed point iterations. Given the state $(W_{n},\Theta_{n})$, the midpoint variables $(\tilde W,\tilde\Theta)$ are found from the first and the third equation in \eqref{MHDmethod1}, and then used to get the updated state $(W_{n+1},\Theta_{n+1})$ by means of the second and the fourth equations in \eqref{MHDmethod1}. Preservation of Casimirs by the method \eqref{MHDmethod1} is guaranteed provided that the midpoint state $(\tilde W,\tilde\Theta)$ is found exactly. In practice, the state $(\tilde W,\tilde\Theta)$ is found up to the tolerance of the fixed point algorithm, and therefore the Casimirs are preserved at least up to that tolerance. In the forthcoming simulations, the tolerance of the fixed point iterations is $10^{-13}$.

Contrary to generic Lie--Poisson integrators, the method \eqref{MHDmethod1} is free of matrix exponentials and therefore scales efficiently as the matrix size grows. The complexity per time step is $O(N^{3})$, see~\citet{CiViMo2023}.

\subsection{The Hazeltine--Zeitlin and CHM--Zeitlin equations}

Before we write the corresponding discretization for Hazeltine's model \eqref{Alfven}, let us note that it is convenient to introduce a new field $q=\omega-\chi$. Then the system \eqref{Alfven} becomes
\begin{equation}
\label{Alfven1}
\left\{
\begin{aligned}
&\dot q=\left\{q,\psi\right\}, \\
&\dot\theta=\left\{\theta,\psi-\alpha\chi\right\},\\
&\dot\chi=\left\{\chi,\psi\right\}+\left\{\theta,j\right\},
\end{aligned}
\right.
\end{equation}
and the corresponding semidirect product Lie--Poisson bracket for \eqref{Alfven1} in terms of the new fields $q,\theta,\chi$ simplifies to
\begin{equation*}
%\label{alfbracket1}
\begin{split}
 \llbracket F,G\rrbracket&=\int\limits_{M}\left[\chi\left\{\frac{\delta F}{\delta\chi},\frac{\delta G}{\delta\chi}\right\}+
 \theta\left(\left\{\frac{\delta F}{\delta\theta},\frac{\delta G}{\delta\chi}\right\}+\left\{\frac{\delta F}{\delta\chi},\frac{\delta G}{\delta\theta}\right\}\right)+q\left\{\frac{\delta F}{\delta q},\frac{\delta G}{\delta q}\right\}\right]\mathrm{d}x\mathrm{d}y.
 \end{split}
\end{equation*}
The \emph{Hazeltine--Zeitlin equations} for \eqref{Alfven1} are then
\begin{equation}
\label{qAlfven}
\left\{
\begin{aligned}
&\dot Q=\frac{1}{\hbar}[Q,P], \\
&\dot\Theta=\frac{1}{\hbar}[\Theta,P-\alpha R],\\
&\dot R=\frac{1}{\hbar}[R,P]+\frac{1}{\hbar}[\Theta,J],
\end{aligned}
\right.
\end{equation}
where $R\in\mathfrak{su}(N)$ is the matrix for the field $\chi$. 

The introduction of the new field $q$ reveals the underlying Lie algebra for the matrix system \eqref{qAlfven}. 
Indeed, the first equation, for the $Q$ matrix, is isospectral, and the last two equations for matrices $\Theta$ and $R$ constitute the Lie--Poisson flow on the dual of the semidirect product Lie algebra, similar to \eqref{qMHDvort}, and all the three equations are coupled through the Hamiltonian function. 
Summarizing, the flow \eqref{qAlfven} is a Lie--Poisson system on the dual $\mathfrak{f}^{*}$ of the Lie algebra $$\mathfrak{f}=\mathfrak{su}(N)\oplus(\mathfrak{su}(N)\ltimes\mathfrak{su}(N)^{*}),$$ with the Casimir invariants
\begin{equation}
\label{qcasAlfven}
\mathcal{C}_{f}^{N}=\frac{4\pi}{N}\mathrm{tr}\left(f(\Theta)\right),\quad 
\mathcal{I}_{g}^{N}=\frac{4\pi}{N}\mathrm{tr}(Rg(\Theta)),\quad
\mathcal{P}_{h}^{N}=\frac{4\pi}{N}\mathrm{tr}(k(Q)), 
\end{equation}
for arbitrary smooth functions $f,g,k$, and the Hamiltonian
\begin{equation}
\label{hamAlfven}
H=\frac{2\pi}{N}\mathrm{tr}\left(W^{\dag}P+\Theta^{\dag} J-\alpha R^{\dag}R\right).
\end{equation}

A Lie--Poisson time integrator for \eqref{qAlfven} is obtained by combining the isospectral integrator for the matrix $Q$, and the magnetic midpoint integrator for the matrices $R$ and $\Theta$, which gives the scheme
\begin{equation}
\label{IsoAlfven}
\begin{aligned}
&\Theta_{n}=\tilde\Theta-\frac{\varepsilon}{2}[\tilde\Theta,\tilde{M}]-\frac{\varepsilon^{2}}{4}\tilde{M}\tilde\Theta\tilde{M}, \\
&\Theta_{n+1}=\Theta_{n}+\varepsilon[\tilde\Theta,\tilde{M}],\\
&Q_{n}=\tilde{Q}-\frac{\varepsilon}{2}[\tilde{Q},\tilde{P}]-\frac{\varepsilon^{2}}{4}\tilde{P}\tilde{Q}\tilde{P}, \\
&Q_{n+1}=Q_{n}+\varepsilon[\tilde{Q},\tilde{P}],\\
&R_{n}=\tilde{R}-\frac{\varepsilon}{2}[\tilde{R},\tilde{M}]-\frac{\varepsilon}{2}[\tilde\Theta,\tilde{J}]-\frac{\varepsilon^{2}}{4}\left(\tilde{M}\tilde{R}\tilde{M}+\tilde{J}\tilde\Theta\tilde{M}+\tilde{M}\tilde\Theta\tilde{J}\right),\\
&R_{n+1}=R_{n}+\varepsilon[\tilde{R},\tilde{M}]+\varepsilon[\tilde\Theta,\tilde{J}],
\end{aligned}
\end{equation}
where $\tilde{M}=\tilde{P}-\alpha\tilde{R}$ and with $\varepsilon=\delta t/\hbar$ as before. 
Similar to the integrator \eqref{MHDmethod1}, the scheme \eqref{IsoAlfven} preserves the Casimirs \eqref{qcasAlfven} and nearly preserves the Hamiltonian \eqref{hamAlfven}.

Finally, we present the \emph{CHM--Zeitlin equation} for the matrix discretization of CHM equations \eqref{CHM}, namely
\begin{equation}
\label{ZeitCHM}
\dot\Sigma=\frac{1}{\hbar}[\Sigma,P],\quad \Sigma=\Delta_{N}P- P/\alpha.
\end{equation}
The Casimir invariants and the Hamiltonian for \eqref{ZeitCHM} are given by
\begin{equation*}
\mathcal{C}_{f}^{N}=\frac{4\pi}{N}\mathrm{tr}(f(\Sigma)),\quad H(\Sigma)=\frac{2\pi}{N}\mathrm{tr}(\Sigma^{\dag}P).
\end{equation*}
In this case, Lie--Poisson time integration is obtained via the \emph{isospectral midpoint method} \citep{ModViv}
\begin{equation*}
\Sigma_{n}=\tilde\Sigma-\frac{\varepsilon}{2}[\tilde\Sigma,\tilde{P}]-\frac{\varepsilon^{2}}{4}\tilde{P}\tilde\Theta\tilde{P},\quad \Sigma_{n+1}=\tilde\Sigma+\varepsilon[\tilde\Sigma,\tilde{P}],\quad \tilde\Sigma=\Delta_{N}\tilde{P}-\tilde{P}/\alpha.
\end{equation*}

\section{Long-time behaviour}\label{sec:long_time_simulations}

In this section, we study, via the matrix discretization methods from the previous section, the long-time behaviour for each of the three reduced magnetized fluid models.

\subsection{Reduced magnetohydrodynamics (RMHD)} 
\label{sect4}
The simulation results for RMHD are obtained by using the method \eqref{MHDmethod1}. 
In all the simulations, the matrix dimension is $N=512$, and the initial condition is prepared such that the fields contain $(l_{max}+1)^{2}$ terms in the spherical harmonics basis, with $l_{max}=10$ and with each coefficient randomly generated from the normal distribution. 
One can observe that the Hamiltonian \eqref{qHamiltMHD} splits into two parts: the kinetic part $(2\pi/N)\mathrm{tr}(W^{\dag}P)$ produced by the vorticity dynamics, and the magnetic part $(2\pi/N)\mathrm{tr}(\Theta^{\dag}J)$ produced by the magnetic potential dynamics.
We generate the initial conditions in such a way that kinetic and magnetic energy terms initially contribute approximately equally to the total energy $H(W,\Theta)$.
The final time for the simulations is $T=559$.

For non-magnetic fluids, hydrodynamical turbulence, which in the ideal setting is modelled by the vorticity transport equation, is known to be fundamentally different in 2-D and 3-D. 
In particular, 2-D hydrodynamical turbulence is characterized by the inverse energy cascade, in contrast to 3-D turbulence which has a forward energy cascade. 
In RMHD, the addition of the Lorentz force to the vorticity advection equation leads to rapid amplification of the vorticity via the formation of vortex filaments. 
This cannot happen in 2-D turbulence, where the vorticity function is advected, so its values cannot change.
From the point-of-view of analysis, the situation for RMHD is more similar to the 3-D Euler equations than to the 2-D Euler equations.
Indeed, the RMHD equations~\eqref{MHDvort} are similar in structure to the axisymmetric 3-D Euler equations \citep{ModinPreston2025}, which exhibit finite-time blow-up of vorticity.
On the other hand, the equilibrium spectra obtained by \citet{FyfeMont1976} suggest an inverse cascade of the mean square magnetic potential.
This inverse cascade of the mean square magnetic potential is believed to be a mechanism leading to the formation of large magnetic eddies.

Let us now see if and how these phenomena are manifested in our simulations.

\begin{figure}
    \begin{minipage}[t]{0.5\textwidth}
    % \captionsetup{width=1\textwidth,labelformat=empty} %
      \centering {$\psi$ field,  $t=0$ \phantom{hejhejhej}}
      \includegraphics[width=\linewidth]{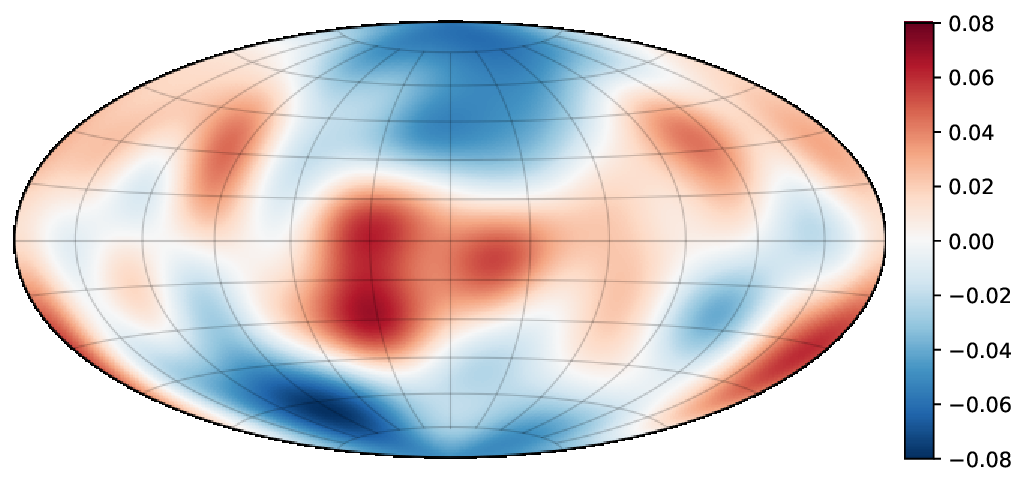} 
    \end{minipage}%
    \hfill % maximize separation between the minipages
    \begin{minipage}[t]{0.5\textwidth}
    % \captionsetup{width=1\textwidth,labelformat=empty}
      \centering {$\theta$ field,  $t=0$ \phantom{hejhejhej}}
      \includegraphics[width=\linewidth]{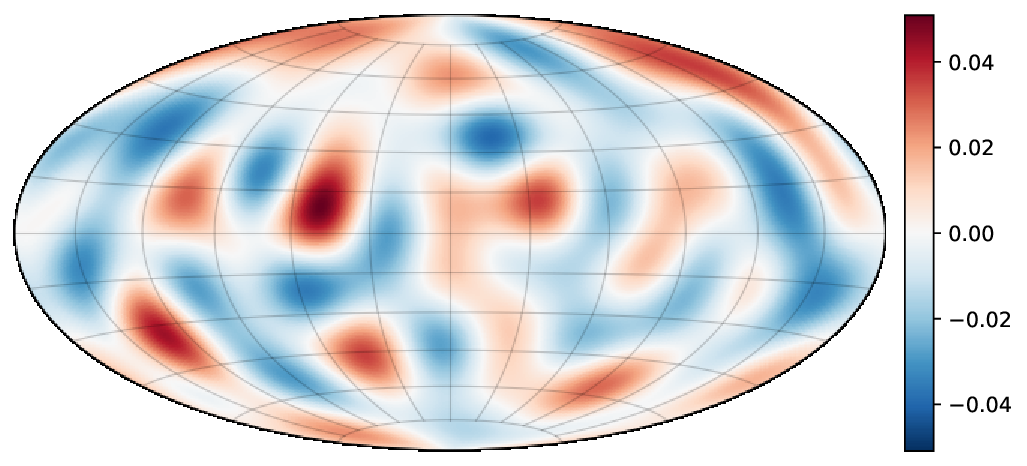} 
    \end{minipage}%
    \hfill % maximize separation between the minipages
    \begin{minipage}[t]{0.5\textwidth}
    % \captionsetup{width=1\textwidth,labelformat=empty}
      \centering {$\psi$ field,  $t=15$ \phantom{hejhejhej}}
      \includegraphics[width=\linewidth]{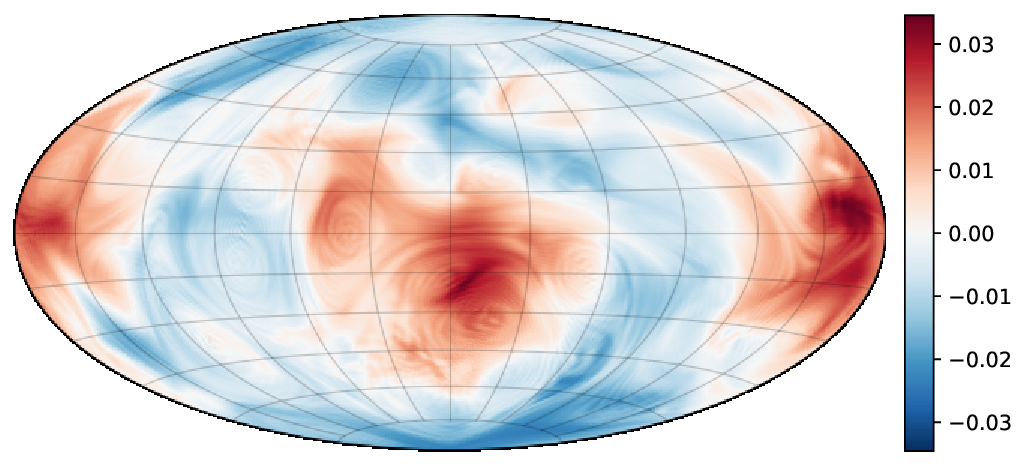} 
    \end{minipage}%
    \hfill % maximize separation between the minipages
    \begin{minipage}[t]{0.5\textwidth}
    % \captionsetup{width=1\textwidth,labelformat=empty}
      \centering {$\theta$ field,  $t=15$ \phantom{hejhejhej}}
      \includegraphics[width=\linewidth]{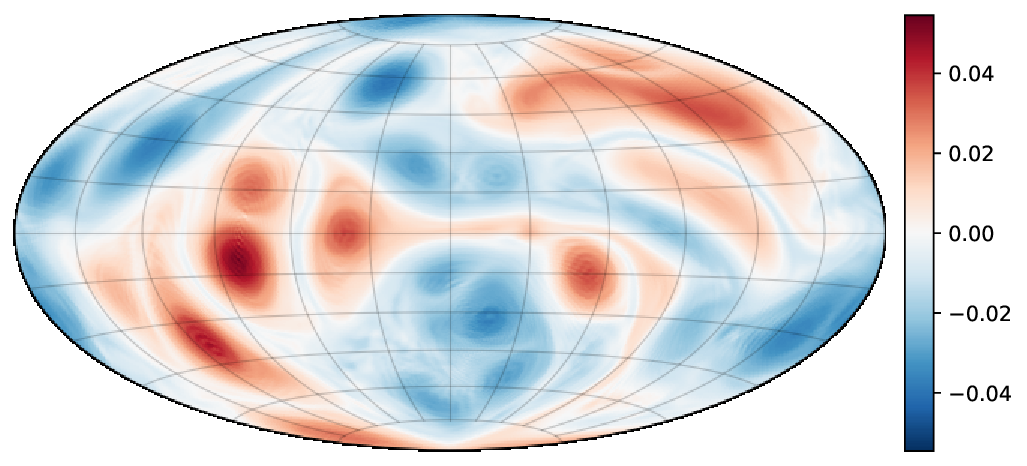} 
    \end{minipage}
    \begin{minipage}[t]{0.5\textwidth}
    % \captionsetup{width=1\textwidth,labelformat=empty} %
      \centering {$\psi$ field,  $t=103$ \phantom{hejhejhej}}
      \includegraphics[width=\linewidth]{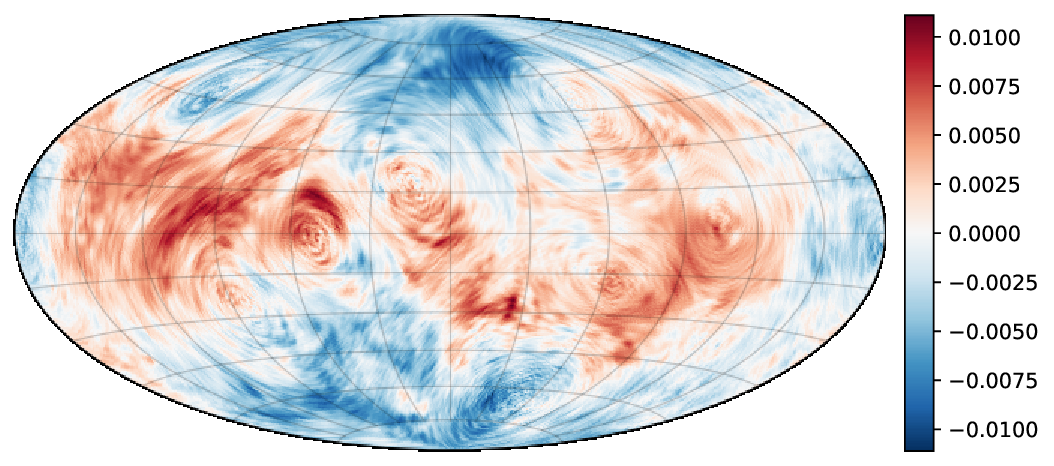} 
    \end{minipage}%
    \hfill % maximize separation between the minipages
    \begin{minipage}[t]{0.5\textwidth}
    % \captionsetup{width=1\textwidth,labelformat=empty}
      \centering {$\theta$ field,  $t=103$ \phantom{hejhejhej}}
      \includegraphics[width=\linewidth]{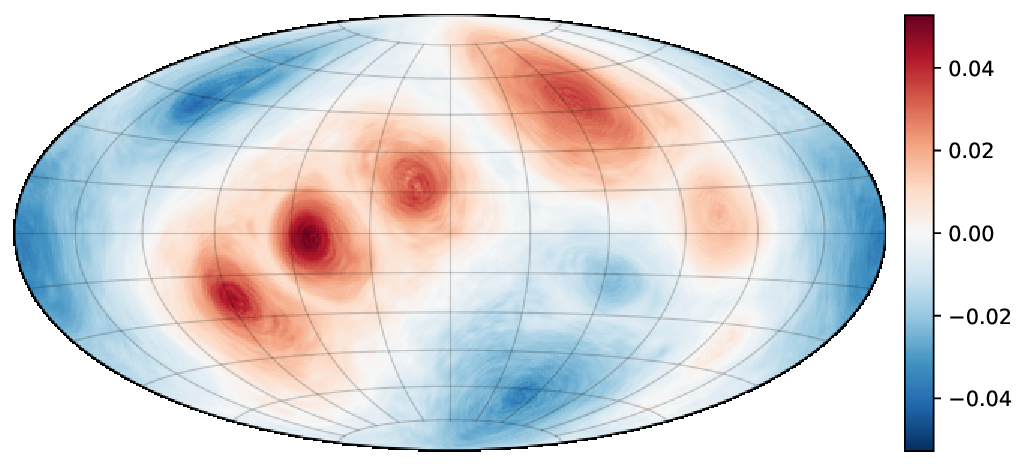} 
    \end{minipage}%
    \hfill % maximize separation between the minipages
    \begin{minipage}[t]{0.5\textwidth}
    % \captionsetup{width=1\textwidth,labelformat=empty}
      \centering {$\psi$ field,  $t=559$ \phantom{hejhejhej}}
      \includegraphics[width=\linewidth]{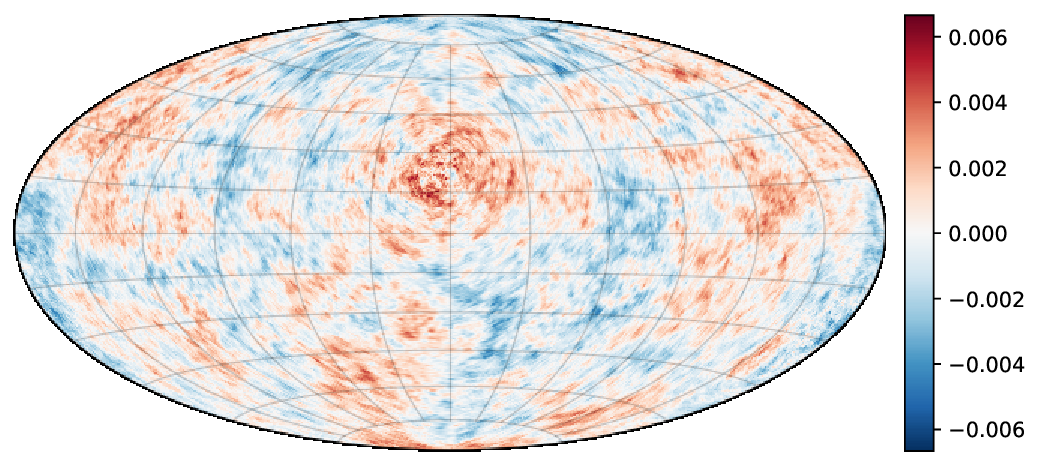} 
    \end{minipage}%
    \hfill % maximize separation between the minipages
    \begin{minipage}[t]{0.5\textwidth}
    % \captionsetup{width=1\textwidth,labelformat=empty}
      \centering {$\theta$ field,  $t=559$ \phantom{hejhejhej}}
      \includegraphics[width=\linewidth]{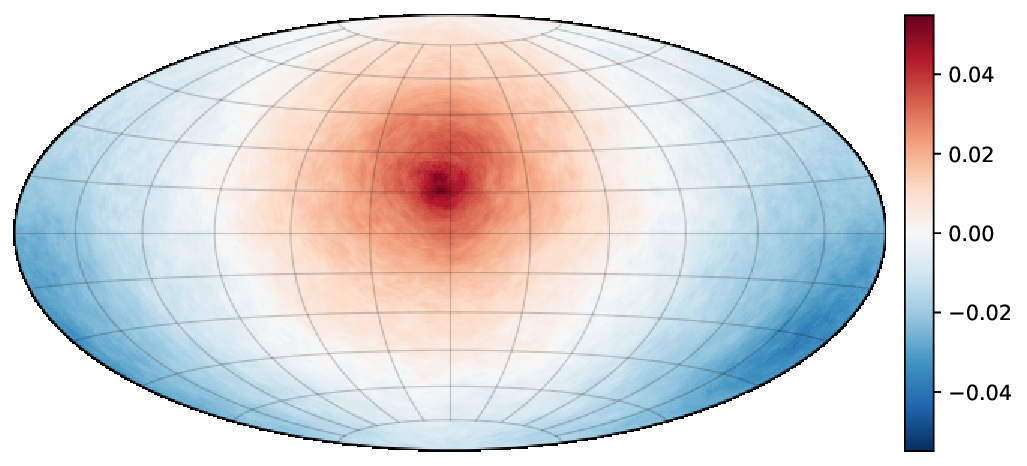} 
    \end{minipage}
\caption{\textbf{RMHD:} Evolution of the velocity stream function $\psi(t)$ (left) and the magnetic potential $\theta(t)$ (right). 
The magnetic potential $\theta$ develops into the dipole configuration through intermediate mixing. 
The stream function $\psi$ does not develop large-scale structures, but one can observe circulations at locations of the magnetic eddies.}
\label{gen-magn-evol-figure}
\end{figure}

In our study, we divide the dynamics into a ``smooth'' or ``slow'' part, represented by the fields $\psi$ and $\theta$, and ``rough'' or ``fast'' part, represented by the fields $\omega$ and $j$.
For the slow part, the evolution of the stream function $\psi$ and the magnetic potential $\theta$ is shown in Figure~\ref{gen-magn-evol-figure}. 
We observe the formation of a magnetic dipole consisting of positive and negative blobs obtained through intermediate mixing. 
The stream function $\psi(t)$ becomes grainy in its long-time behaviour, with no sign of formation of large scale structures, albeit traces of the large-scale structure in $\theta(t)$ are visible. 
In particular, vortex formations (i.e., localized swirling) for $\psi(t)$ and $\theta(t)$ are observed approximately at the same locations.

\begin{figure} %[ht!]
% \centering
% \includegraphics[scale=0.5]{L_infinity_norm}
%\includegraphics{vorticity_growth.pdf}
\includegraphics{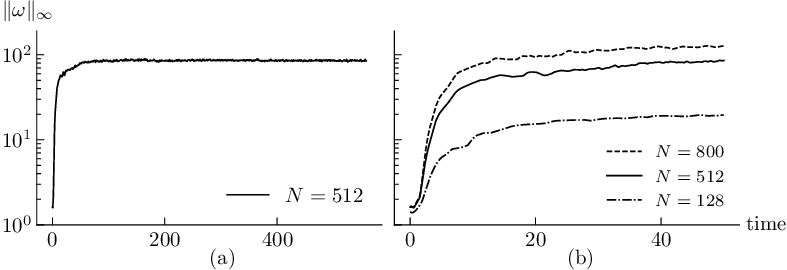}
\caption{\textbf{RMHD:} Evolution of the vorticity field supremum norm $\lVert \omega\rVert_\infty$. % $\omega(t)$ and the current density $j(t)$.
% \\ 
(a) For a simulation with spatial resolution $N=512$, the value initially grows rapidly and then reach a plateau at about $t=100$.
% After a rapid initial growth, the supremum reaches saturation at about $t=100$.
% \\
(b) For simulations with the same initial data, the plateau is larger in magnitude for higher spatial resolution.
This indicates that the value grows indefinitely as $N\to \infty$.
% This indicates that the RMHD equations may exhibit finite time blow-up of solutions.
}
\label{linfnorms}
\end{figure}

\begin{figure} %[ht]
    \begin{minipage}[t]{0.5\textwidth}
    % \captionsetup{width=1\textwidth,labelformat=empty} %
      \centering {$\omega$ field,  $t=0$ \phantom{hejhejh}}
      \includegraphics[width=\linewidth]{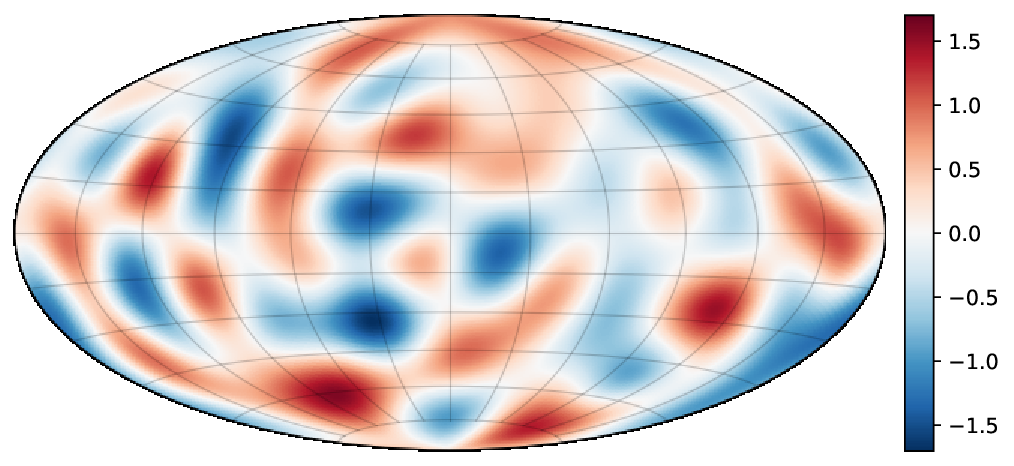} 
    \end{minipage}%
    \hfill % maximize separation between the minipages
    \begin{minipage}[t]{0.5\textwidth}
    % \captionsetup{width=1\textwidth,labelformat=empty}
      \centering {$j$ field,  $t=0$ \phantom{hejhejh}}
      \includegraphics[width=\linewidth]{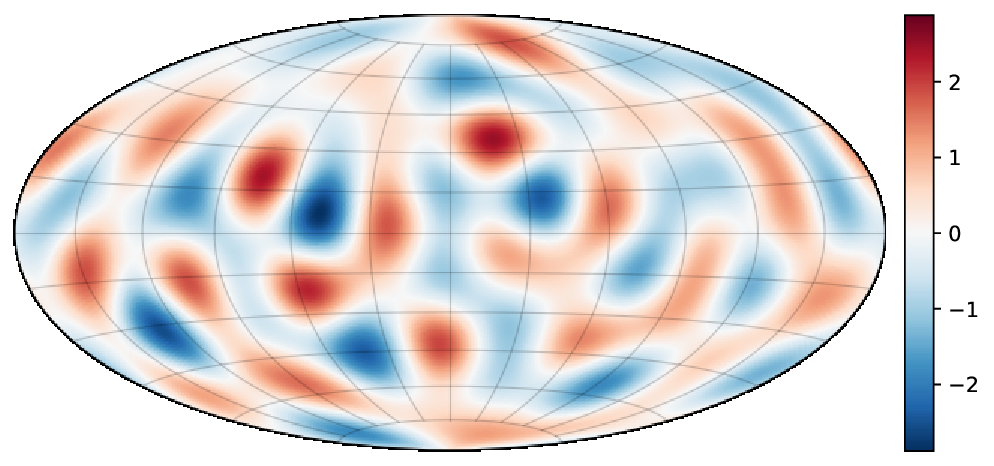} 
    \end{minipage}%
    \hfill % maximize separation between the minipages
    \begin{minipage}[t]{0.5\textwidth}
    % \captionsetup{width=1\textwidth,labelformat=empty}
      \centering {$\omega$ field,  $t=1.2$ \phantom{hejhejh}}
      \includegraphics[width=\linewidth]{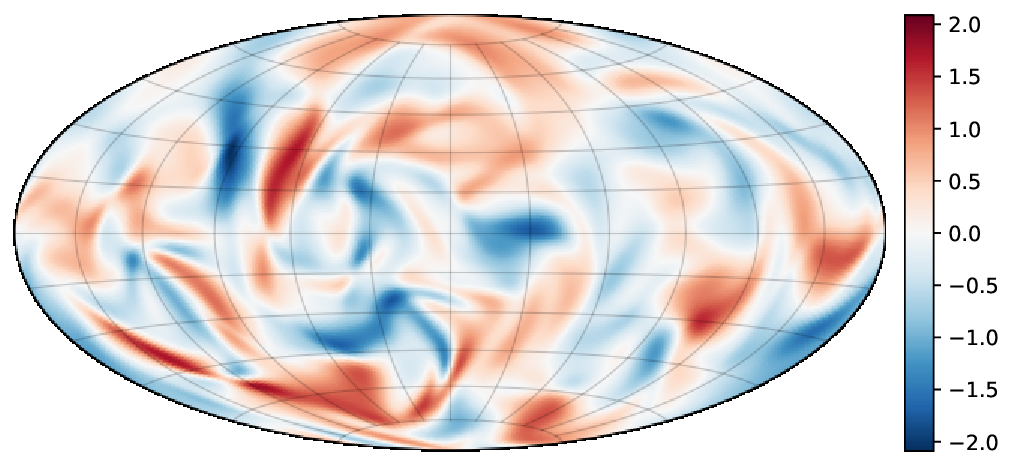} 
    \end{minipage}%
    \hfill % maximize separation between the minipages
    \begin{minipage}[t]{0.5\textwidth}
    % \captionsetup{width=1\textwidth,labelformat=empty}
      \centering {$j$ field,  $t=1.2$ \phantom{hejhejh}}
      \includegraphics[width=\linewidth]{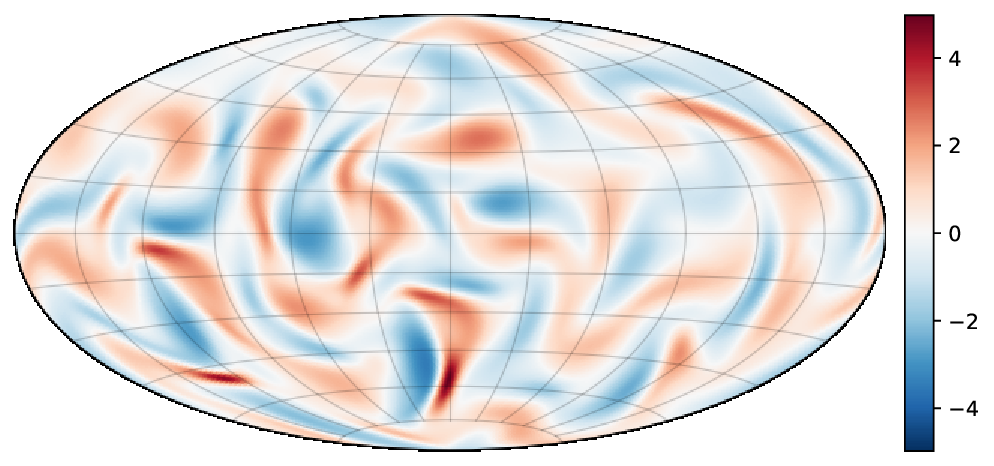} 
    \end{minipage}
    \begin{minipage}[t]{0.5\textwidth}
    % \captionsetup{width=1\textwidth,labelformat=empty} %
      \centering {$\omega$ field,  $t=2$ \phantom{hejhejh}}
      \includegraphics[width=\linewidth]{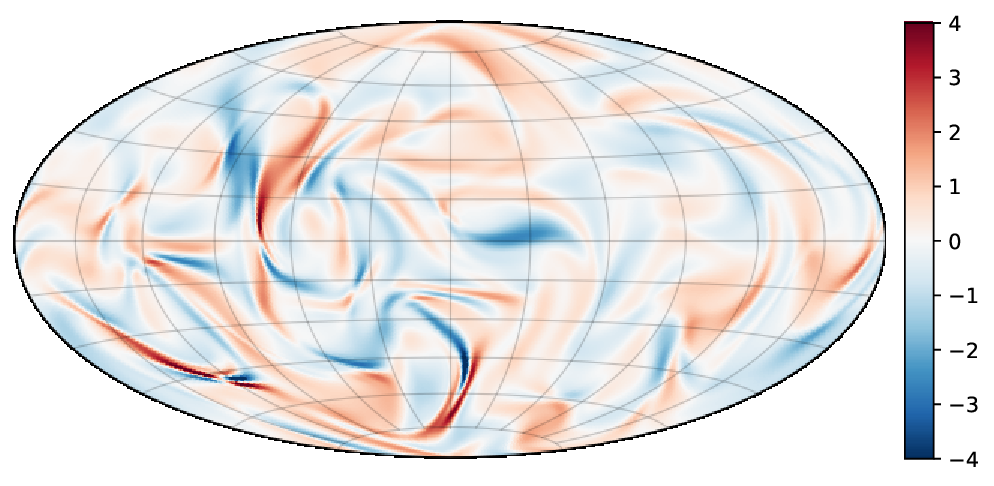} 
    \end{minipage}%
    \hfill % maximize separation between the minipages
    \begin{minipage}[t]{0.5\textwidth}
    % \captionsetup{width=1\textwidth,labelformat=empty}
      \centering {$j$ field,  $t=2$ \phantom{hejhejh}}
      \includegraphics[width=\linewidth]{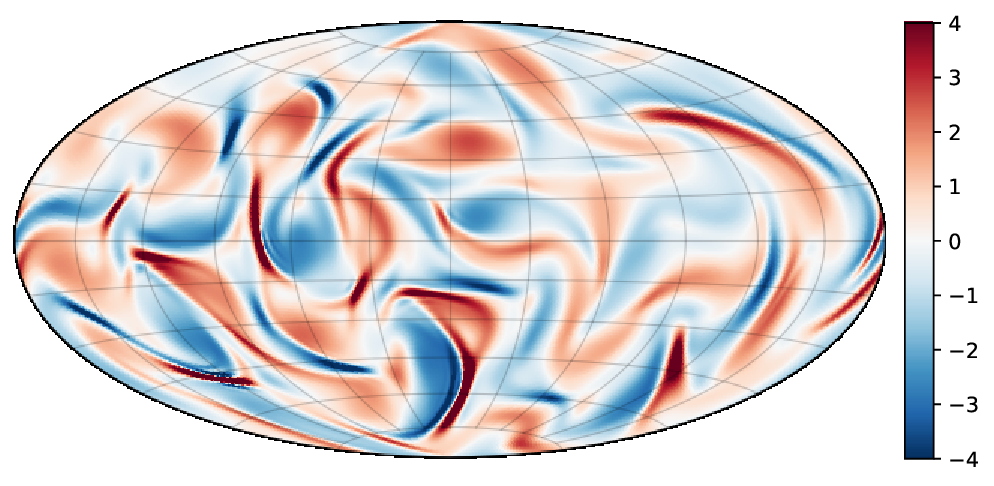} 
    \end{minipage}%
    \hfill % maximize separation between the minipages
    \begin{minipage}[t]{0.5\textwidth}
    % \captionsetup{width=1\textwidth,labelformat=empty}
      \centering {$\omega$ field,  $t=559$ \phantom{hejhejh}}
      \includegraphics[width=\linewidth]{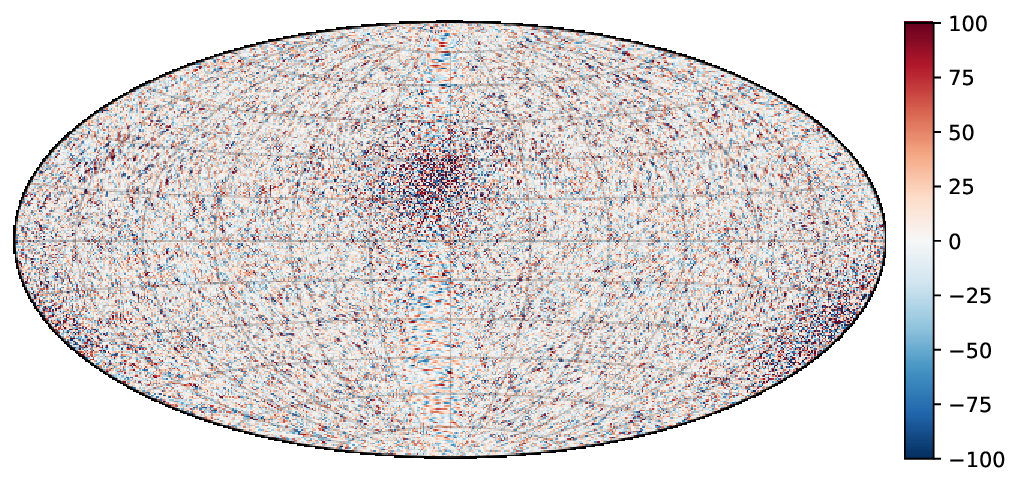} 
    \end{minipage}%
    \hfill % maximize separation between the minipages
    \begin{minipage}[t]{0.5\textwidth}
    % \captionsetup{width=1\textwidth,labelformat=empty}
      \centering {$j$ field,  $t=559$ \phantom{hejhejh}}
      \includegraphics[width=\linewidth]{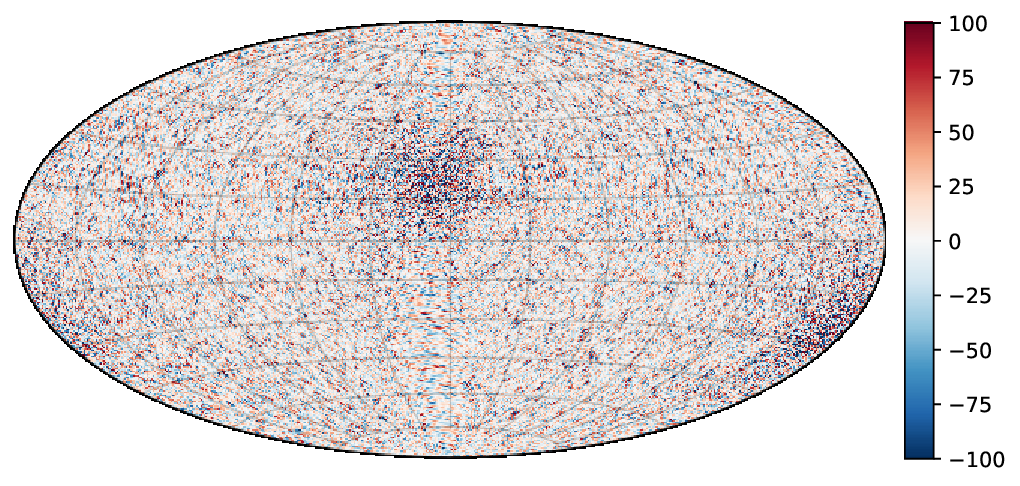} 
    \end{minipage}
\caption{\textbf{RMHD:} Evolution of the vorticity $\omega(t)$ (left) and the current density $j(t)$ (right). 
Vorticity and current density islands resemble each other. Both fields $\omega$ and $j$ are significantly amplified, which makes the dynamics resolved only for relatively short times.}
\label{gen-vort-current-evol-figure}
\end{figure}

For the fast part, the evolution of the fields $\omega(t)$ and $j(t)$ is shown in Figure~\ref{gen-vort-current-evol-figure}. 
We see the formation of filaments in both $\omega$ and $j$.
These are well-resolved only for a short while, up until about $t=2$.
After that, they are too narrow and too entangled to be resolved.
At the final time $t=559$, we obtained a noisy result with large vorticity values.
One should interpret this noise as the truncation of an extremely delicate filament structure.
Notice in Figure~\ref{linfnorms}(a) that the supremum norm of $\omega$ rapidly grows from about $2$ to about $100$ (the current density field $j$ shadows $\omega$, with rapid growth from about $2$ to about $100$). 
These values settle at larger magnitudes if a higher spatial resolution is used, which is shown in Figure~\ref{linfnorms}(b), where we present the evolution of the supremum norm of the vorticity field $\omega$ for resolutions $N=128$, $N=512$, and $N=800$. 
This behaviour is similar to the axisymmetric 3-D Euler equations, where numerical simulations via matrix hydrodynamics also show growth of the supremum norm of vorticity for growing spatial resolution \citep{ModinPreston2025}.
As mentioned, the axisymmetric 3-D Euler equations are known to have solutions that lead to blow-up of the vorticity in finite time.
Whether the RMHD equations exhibit finite time blow-up is an open mathematical analysis problem.

% \begin{figure} %[ht!]
% \centering
% \includegraphics[scale=0.5]{L_infinity_norm_all}
% \caption{\textbf{RMHD:} Evolution of the supremum norm of the vorticity field $\omega(t)$. The supremum norm settles at larger magnitudes for higher spatial resolution. This indicates that the RMHD equations may exhibit finite time blow-up of solutions.
% }
% \label{linfnorms_all}
% \end{figure}

\begin{figure}
\centering
\includegraphics{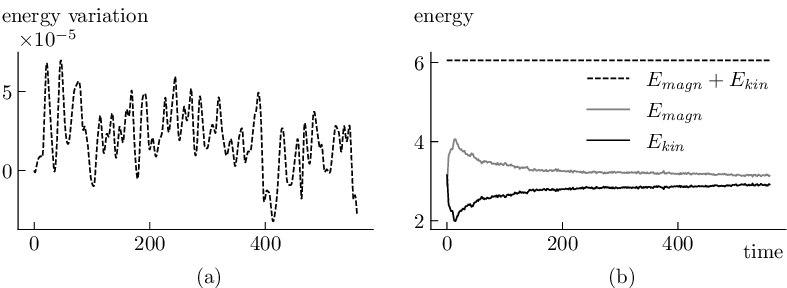}
\caption{\textbf{RMHD:} (a) Total energy variation, and (b) kinetic $E_{kin}$ and magnetic $E_{magn}$ energy evolution over time. 
The total energy $E_{magn} + E_{kin}$ is conserved up to a relative error of about $10^{-5}$. 
The magnetic and kinetic energy components are redistributed with a tendency towards equipartition.}
\label{hamilt-figure}
\end{figure}

In Figure~\ref{hamilt-figure}, we demonstrate the evolution of the total energy, as well as its kinetic $E_{kin}$ and magnetic $E_{magn}$ components. While the total energy is conserved, its kinetic and magnetic parts redistribute in such a way that magnetic energy initially grows in time, reaches its maximum, and then decays while dominating over the kinetic energy. Kinetic energy has the opposite dynamics. In the long term, there is a tendency of equipartition between kinetic and magnetic energies. 

\begin{figure}
  \includegraphics{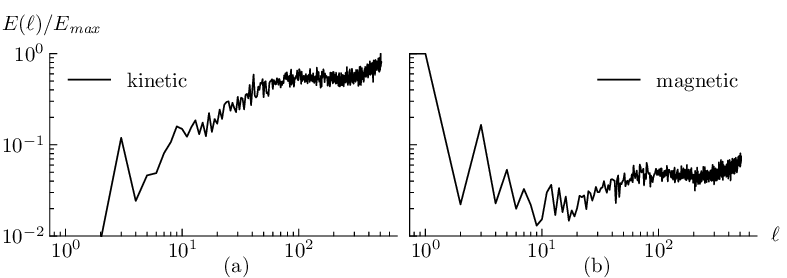}
    % \begin{minipage}[t]{0.5\textwidth}
    % % \captionsetup{width=1\textwidth,labelformat=empty} %
    %   \includegraphics[width=\linewidth]{loglog_kin_en_spec_MHD} 
    %   %\centering {Energy spectrum at initial time $t=0$}
    % \end{minipage}%
    % \hfill % maximize separation between the minipages
    % \begin{minipage}[t]{0.5\textwidth}
    % % \captionsetup{width=1\textwidth,labelformat=empty}
    %   \includegraphics[width=\linewidth]{loglog_magn_en_spec_MHD} 
    %   %\caption*{Energy spectrum at final time $t=633$}
    % \end{minipage}%
  \caption{\textbf{RMHD:} Kinetic (a) and magnetic (b) energy spectra at the final time $t=559$. 
  A major part of the kinetic energy is concentrated at high frequencies, indicating a forward kinetic energy cascade. 
  The magnetic energy is mostly concentrated at lower frequencies and thereafter homogeneously distributed over the higher frequencies, indicating a backward cascade of magnetic energy; the effect is further emphasized in the spectrum of the mean square magnetic potential, shown in Figure~\ref{spec-MHD-pot}.}
\label{spec-MHD}
\end{figure}

The normalized spectra for magnetic and kinetic energies at the final time $t=559$ are presented in Figure~\ref{spec-MHD}. 
As predicted by the statistical theory, the kinetic energy experiences a direct cascade, as most of it is concentrated in the modes of high $l$. 
The magnetic energy, on the contrary, is localized at lower frequencies, which is reflected by larger magnitudes in its spectrum at low frequencies. This also reflects simulations in Figure~\ref{gen-magn-evol-figure} showing the formation of the large-scale magnetic dipole. Both kinetic and magnetic energies are uniformly distributed over higher frequencies, as seen in Figure~\ref{spec-MHD}.

\begin{figure}
  \includegraphics{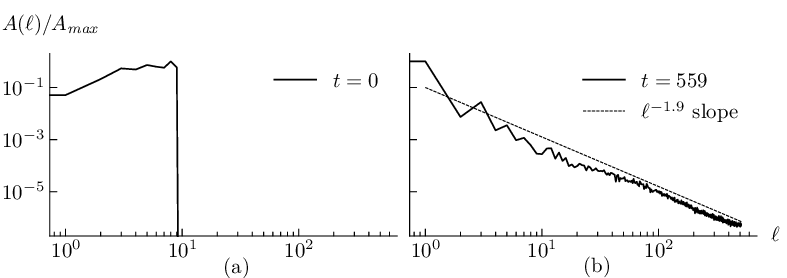}
    % \begin{minipage}[t]{0.52\textwidth}
    % % \captionsetup{width=1\textwidth,labelformat=empty} %
    %   \includegraphics[width=\linewidth]{magn_potential_spec_init} 
    %   %\caption*{Energy spectrum at initial time $t=0$}
    % \end{minipage}%
    % % \hfill % maximize separation between the minipages
    % \begin{minipage}[t]{0.52\textwidth}
    % % \captionsetup{width=1\textwidth,labelformat=empty}
    %   \includegraphics[width=\linewidth]{magn_potential_spec_final} 
    %   %\caption*{Energy spectrum at final time $t=633$}
    % \end{minipage}%
\caption{\textbf{RMHD:} Mean square of magnetic potential $A$ spectrum at (a) initial time $t=0$, and (b) final time $t=559$. 
An inverse cascade of $A$ is observed with the approximate scaling $\ell^{-1.9}$.}
\label{spec-MHD-pot}
\end{figure}

Finally, we present the spectrum for the mean-square magnetic potential $A=\int_{S^{2}}\theta^{2}\mathrm{d}x\mathrm{d}y$ in Figure~\ref{spec-MHD-pot}, which clearly indicates its inverse cascade with the slope approximately $-2$.

\subsection{The Charney--Hasegawa--Mima (CHM) model}
\begin{figure}
    \begin{minipage}[t]{0.5\textwidth}
    % % \captionsetup{width=1\textwidth,labelformat=empty} %
      % \caption*{$\sigma$ field,  $t=0$ \phantom{hejhe}}
      \centering $\sigma$ field,  $t=0$ \phantom{hejhe} \\
      \includegraphics[width=\linewidth]{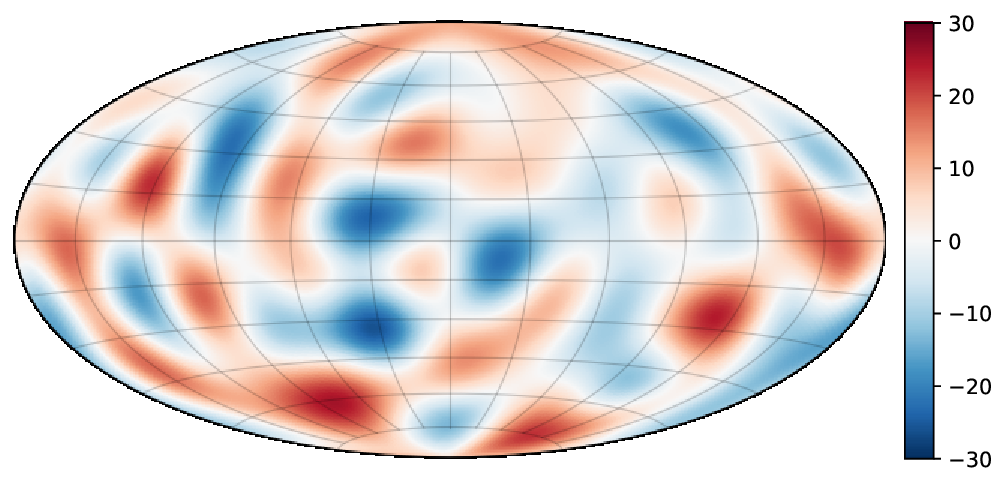} 
    \end{minipage}%
    \hfill % maximize separation between the minipages
    \begin{minipage}[t]{0.5\textwidth}
    % \captionsetup{width=1\textwidth,labelformat=empty}
      \centering {$\sigma$ field,  $t=2.5$ \phantom{hejhe}}
      \includegraphics[width=\linewidth]{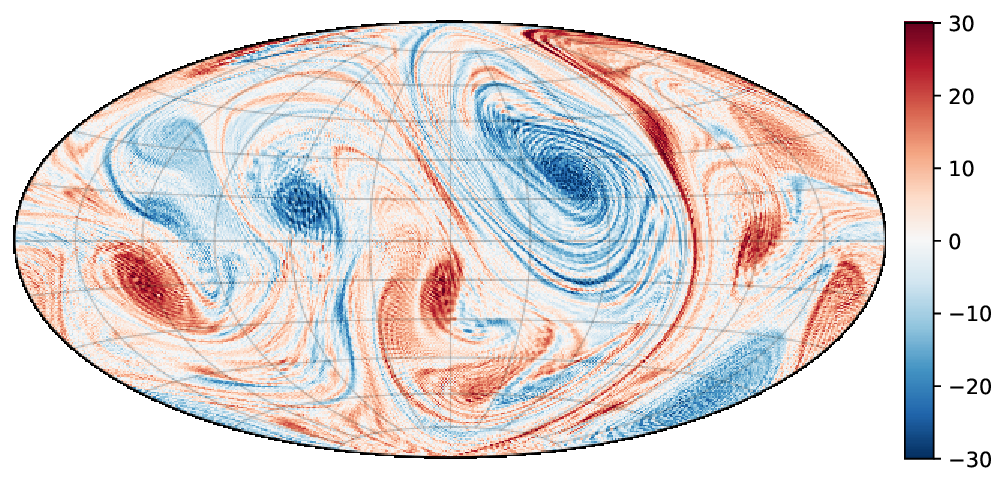} 
    \end{minipage}%
    \hfill % maximize separation between the minipages
    %\begin{comment}
    \begin{minipage}[t]{0.5\textwidth}
    % \captionsetup{width=1\textwidth,labelformat=empty}
      \centering {$\sigma$ field,  $t=4$ \phantom{hejhe}}
      \includegraphics[width=\linewidth]{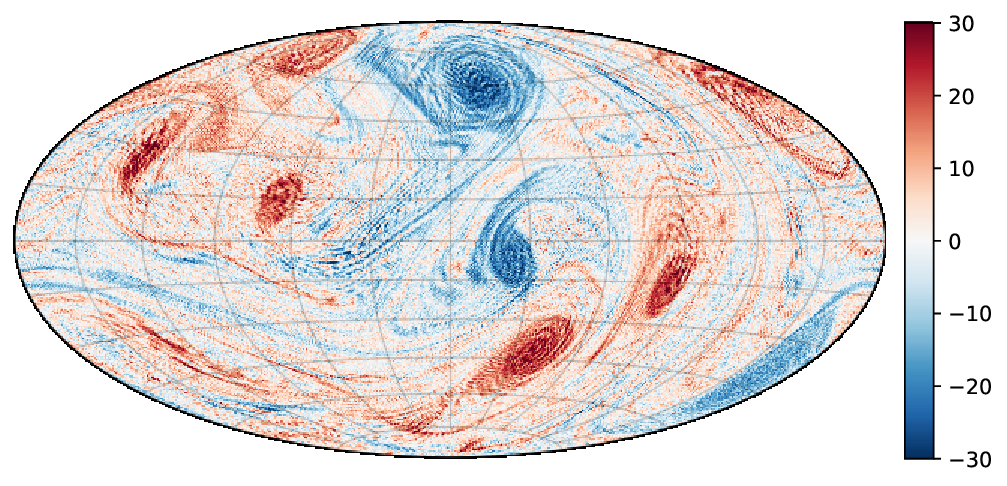} 
    \end{minipage}%
    \hfill % maximize separation between the minipages
    \begin{minipage}[t]{0.5\textwidth}
    % \captionsetup{width=1\textwidth,labelformat=empty}
      \centering {$\sigma$ field,  $t=6$ \phantom{hejhe}}
      \includegraphics[width=\linewidth]{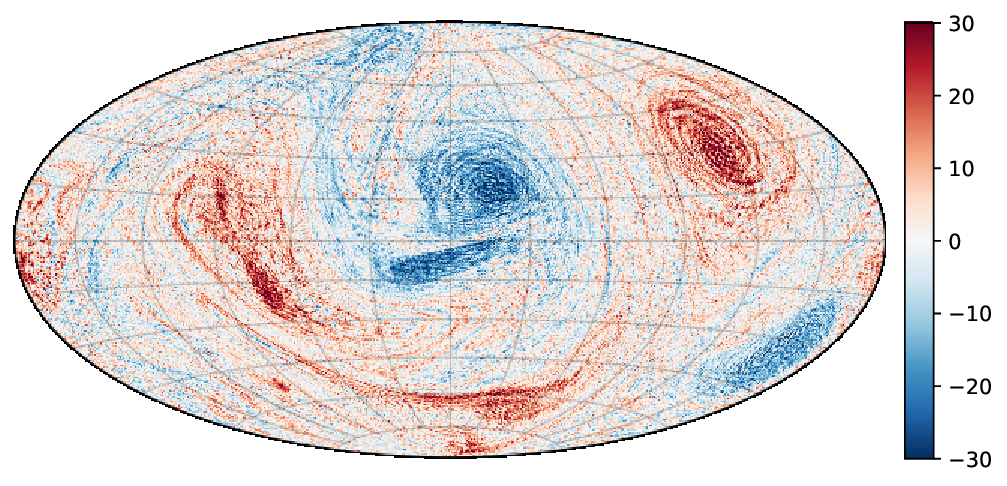} 
    \end{minipage}
    %\end{comment}
    \begin{minipage}[t]{0.5\textwidth}
    % \captionsetup{width=1\textwidth,labelformat=empty}
      \centering {$\sigma$ field,  $t=9.5$ \phantom{hejhe}}
      \includegraphics[width=\linewidth]{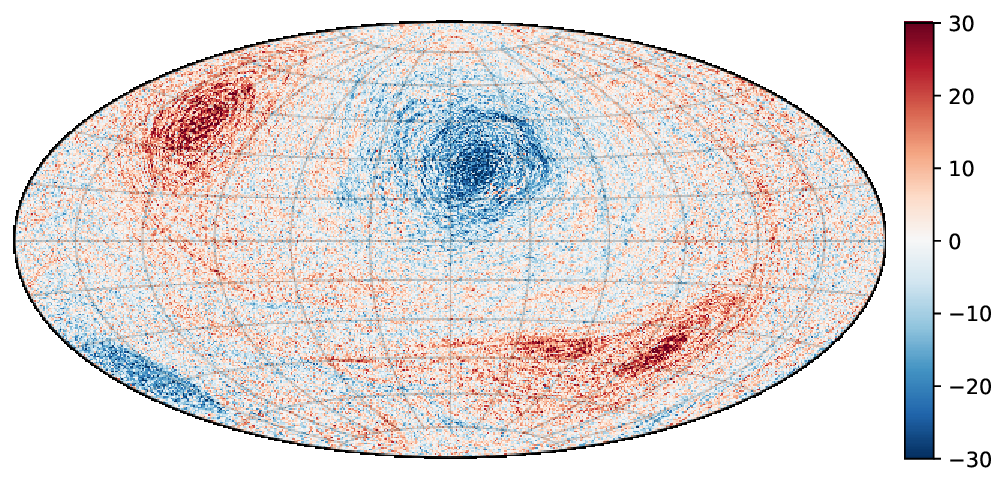} 
    \end{minipage}%
    \hfill % maximize separation between the minipages
    \begin{minipage}[t]{0.5\textwidth}
    % \captionsetup{width=1\textwidth,labelformat=empty}
      \centering {$\sigma$ field,  $t=76$ \phantom{hejhe}}
      \includegraphics[width=\linewidth]{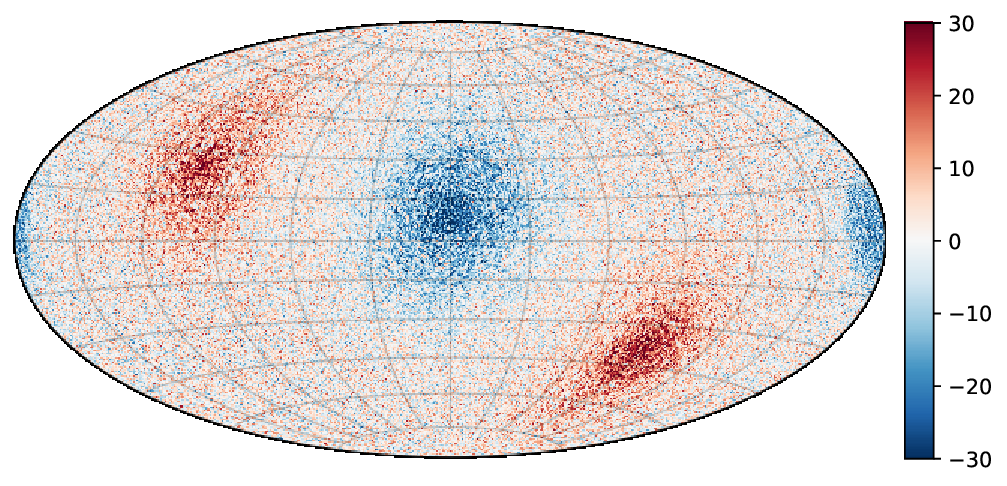} 
    \end{minipage}
\caption{\textbf{CHM:} Evolution of the potential vorticity field $\sigma(t)$. Smooth randomly generated initial distribution evolves into four vortex blob configuration (two positive and two negative) involved in a quasi-periodic motion.
}
\label{CHM_sim}
\end{figure}

For the CHM model simulation, the length scale parameter is chosen to be $\alpha=1/2$. In Figure~\ref{CHM_sim}, we present the evolutions of the potential vorticity field $\sigma(t)$. The spatial resolution parameter $N=512$, and the final simulation time is $t=76$, by which the system reaches statistically equilibrium configuration consisting of two positive and two negative vortex blobs in a quasi-periodic motion. We observe the formation of the four vortex condensates regardless of the value of the total angular momentum, contrary to the Euler equations, where the four vortex blob configuration is reached from initial states having trivial total angular momentum. The mixing phase of merging small vortices of the same sign into larger vortices continues until the time approximately $t=10.5$, when the four blob configuration is reached. 
After that, no further mixing occurs. 

\begin{figure}
\centering
\includegraphics{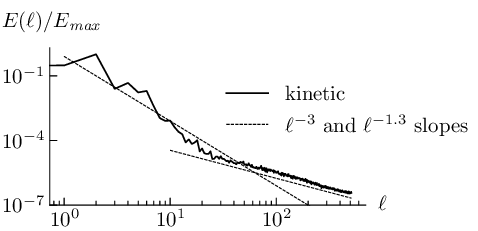}
\caption{\textbf{CHM:} Kinetic energy spectrum of the final state at $t=76$. The spectrum has a broken line shape with the scaling $l^{-3}$ for the low frequency part, and $l^{-1.3}$ for the high frequency part.}
\label{energy_sepctrum_CHM}
\end{figure}

The kinetic energy spectrum is shown in Figure~\ref{energy_sepctrum_CHM}, which indicates the presence of the inverse cascade with the broken line shape of the slope.

\subsection{Hazeltine's model}
\label{sect5}
For the Hazeltine model simulation, we choose the length scale parameter $\alpha=1/2$, and prepare the initial vorticity distribution $\omega_{0}$ with vanishing total angular momentum. 
The motivation for such a choice comes from the long-time simulations of the incompressible Euler equations, where the dynamics settles at the four-blob configuration in the case of the vanishing total angular momentum; see, e.g., \citep{ModViv1}, where it is conjectured that vorticity mixing continues until the system reaches nearly integrable behaviour and remains close to it in the KAM sense. 
As in the previous simulations of the RMHD system, we choose initial condition with equal contribution to the total energy from all the fields.
The spatial resolution is again $N=512$ and the total simulation time is set to $t=562$.

% The chosen spatial resolution $N=512$ allows to resolve the dynamics on a much longer time range, which for this case was chosen to be $t=562$. 
\begin{figure}
    \begin{minipage}[t]{0.5\textwidth}
    % \captionsetup{width=1\textwidth,labelformat=empty} %
    \centering {$\omega$ field,  $t=0$ \phantom{hejh}}
      \includegraphics[width=\linewidth]{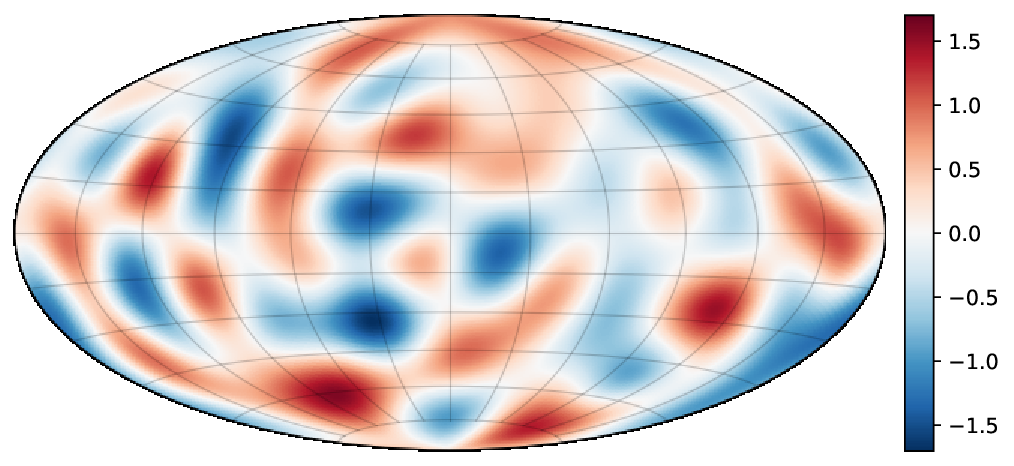} 
    \end{minipage}%
    \hfill % maximize separation between the minipages
    \begin{minipage}[t]{0.5\textwidth}
    % \captionsetup{width=1\textwidth,labelformat=empty}
    \centering {$\omega$ field,  $t=36$ \phantom{hejh}}
      \includegraphics[width=\linewidth]{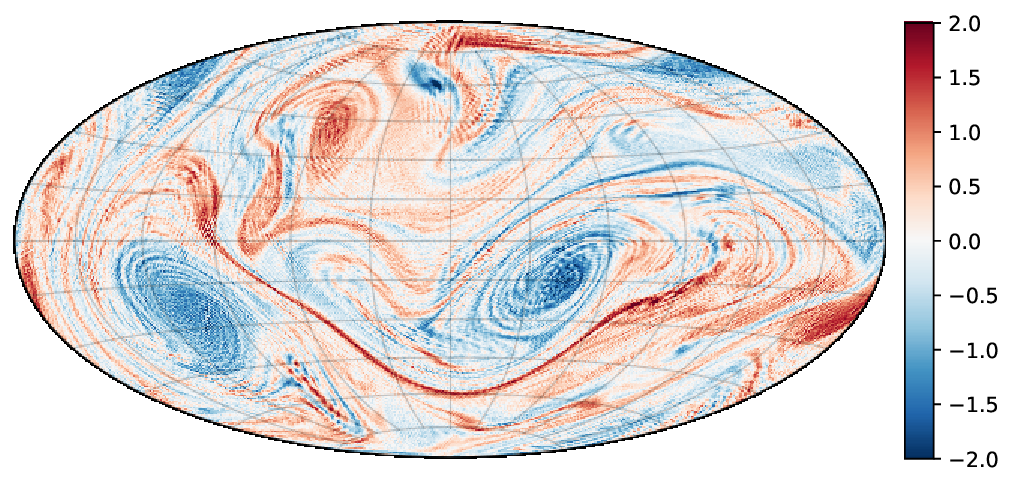} 
      
    \end{minipage}%
    \hfill % maximize separation between the minipages
    \begin{minipage}[t]{0.5\textwidth}
    % \captionsetup{width=1\textwidth,labelformat=empty}
    \centering {$\omega$ field,  $t=120$ \phantom{hejh}}
      \includegraphics[width=\linewidth]{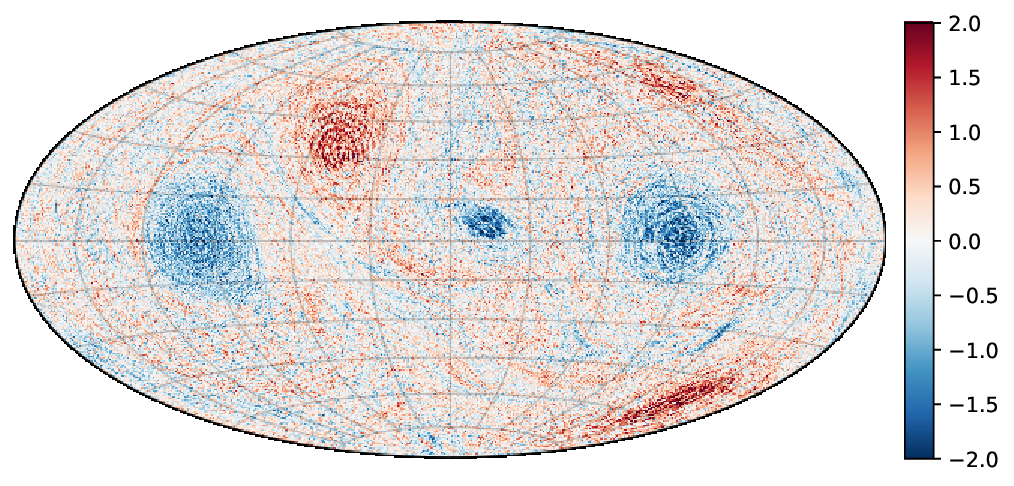} 
      
    \end{minipage}%
    \hfill % maximize separation between the minipages
    \begin{minipage}[t]{0.5\textwidth}
    % \captionsetup{width=1\textwidth,labelformat=empty}
    \centering {$\omega$ field,  $t=562$ \phantom{hejh}}
      \includegraphics[width=\linewidth]{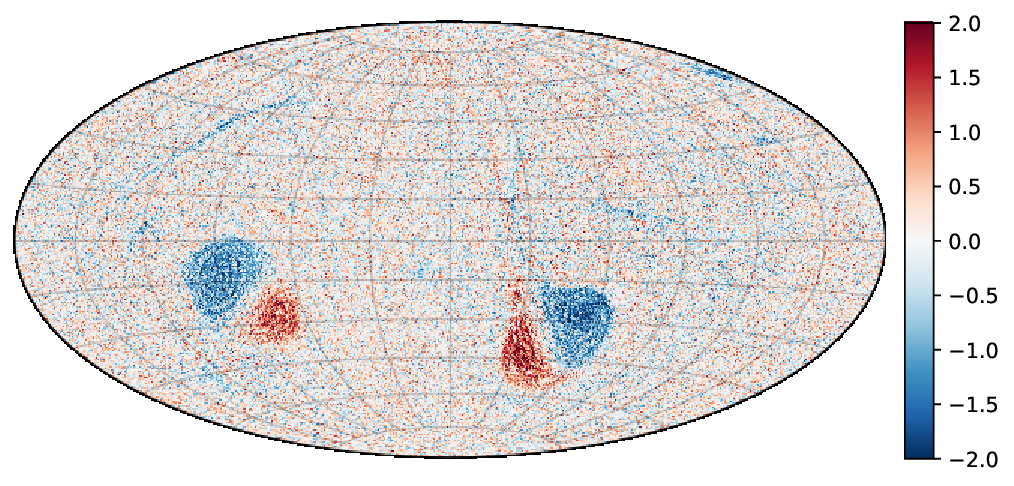} 
      
    \end{minipage}
\caption{\textbf{Hazeltine:} Evolution of the vorticity $\omega(t)$ field. 
Smooth randomly generated initial vorticity distribution evolves into vortex blob configuration on the small scale background noise.}
\label{alf_vort_dynamics}
\end{figure}

As seen in Figure~\ref{alf_vort_dynamics}, the vorticity dynamics in the Hazeltine model is significantly different from that of RMHD. 
Indeed, the vorticity evolution resembles that of the 2-D Euler. 
Namely, one observes the formation of four large scale vortex blobs, two positive and two negative, through intermediate mixing. 
Notice, in particular, that the vorticity values are almost conserved, which is unexpected since the $\omega$ is not transported in Hazeltine's model.
The $\chi$ field seems to compensate the amplification of the vorticity $\omega$ taking place in RMHD dynamics. On the other hand, the blob formation is different from that in the Euler dynamics and the CHM dynamics. 
Indeed, the blobs travel ``in dipole pairs'', where the vorticity dipole is formed by a negative and a positive vortex blob, without however further mixing. In the Euler dynamics, on the contrary, blobs of different sign tend to repel each other.
This behaviour indicates that the extension of RMHD by the density variation field $\chi$ makes the plasma turbulence more alike to the 2-D hydrodynamical turbulence.

\begin{figure}
\centering
\includegraphics{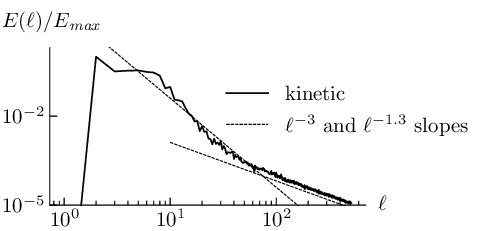}
\caption{\textbf{Hazeltine:} Kinetic energy spectrum of the final state at $t=562$. The spectrum has a broken line shape with the scaling $l^{-3}$ for the low frequency part, and $l^{-1.3}$ for the high frequency part.}
\label{kin_en_spec_alf}
\end{figure}

Indeed, we underline that the vorticity field dynamics in Hazeltine's model has a clear scale separation, which is absent in the long-time RMHD dynamics for the same space resolution: large-scale vortex blobs slowly travel on the small scale rapidly evolving background. 
This is also seen from the kinetic energy spectrum in Figure~\ref{kin_en_spec_alf}. 
The low frequency part of the spectrum, which corresponds to large spatial scales has the slope $l^{-3}$, and the high frequency part corresponding to small spatial scales behaves as $l^{-1.3}$. 
The slopes are in agreement with the spectral properties of the Euler flows, see \citep{ModViv2}, where one observes a similar scale separation with the slopes $l^{-1}$ for small scales and $l^{-3}$ for large scales.

\begin{figure}
    \begin{minipage}[t]{0.5\textwidth}
    % \captionsetup{width=1\textwidth,labelformat=empty} %
      \centering {$\psi$ field,  $t=0$ \phantom{hejhejhej1}}
      \includegraphics[width=\linewidth]{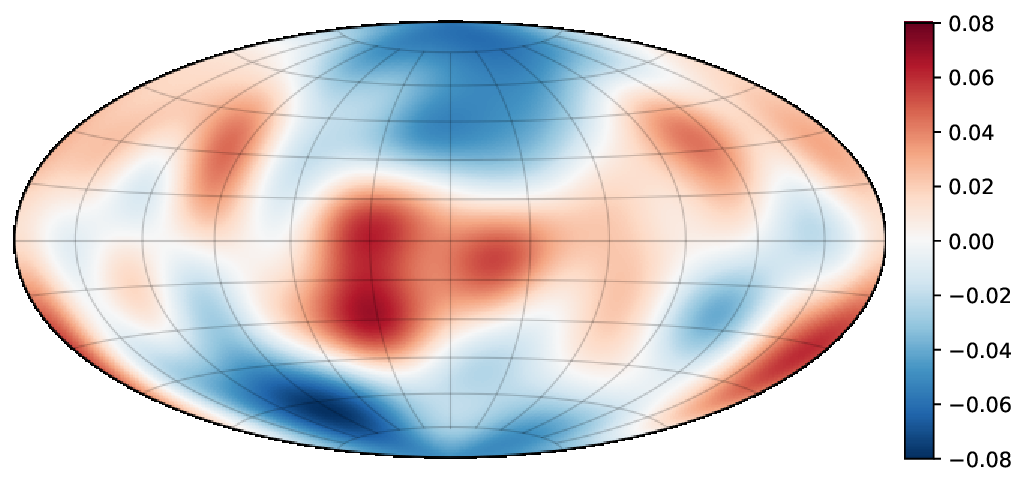} 
    \end{minipage}%
    \hfill % maximize separation between the minipages
    \begin{minipage}[t]{0.5\textwidth}
    % \captionsetup{width=1\textwidth,labelformat=empty}
      \centering {$\theta$ field,  $t=0$ \phantom{hejhejhej1}}
      \includegraphics[width=\linewidth]{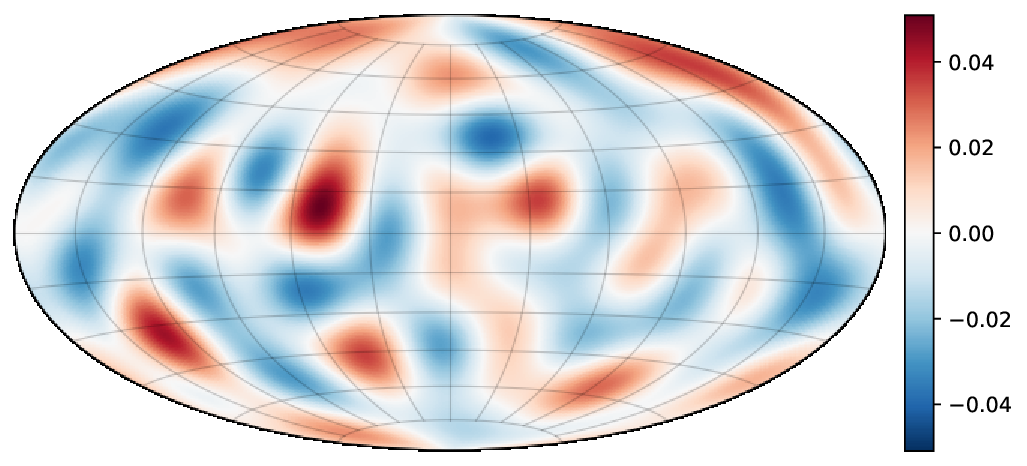} 
    \end{minipage}%
    \hfill % maximize separation between the minipages
    \begin{minipage}[t]{0.5\textwidth}
    % \captionsetup{width=1\textwidth,labelformat=empty}
      \centering {$\psi$ field,  $t=8$ \phantom{hejhejhej1}}
      \includegraphics[width=\linewidth]{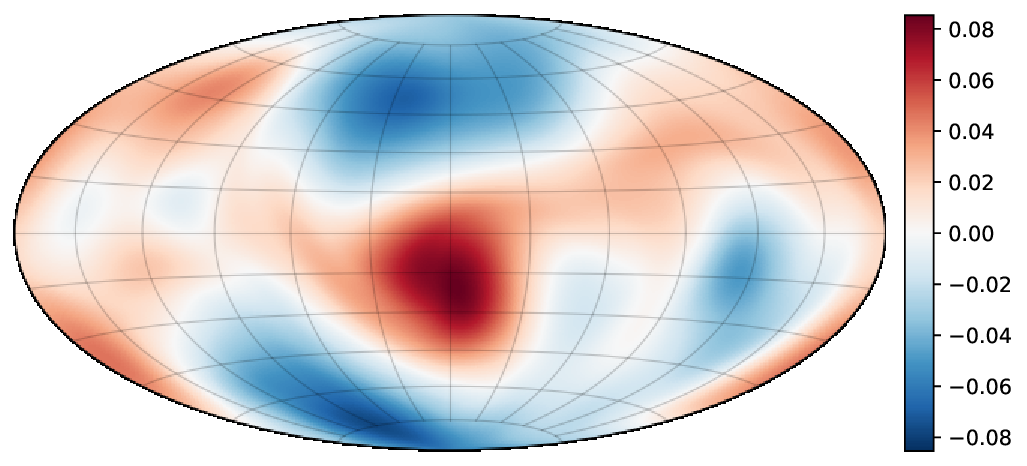} 
    \end{minipage}%
    \hfill % maximize separation between the minipages
    \begin{minipage}[t]{0.5\textwidth}
    % \captionsetup{width=1\textwidth,labelformat=empty}
      \centering {$\theta$ field,  $t=8$ \phantom{hejhejhej1}}
      \includegraphics[width=\linewidth]{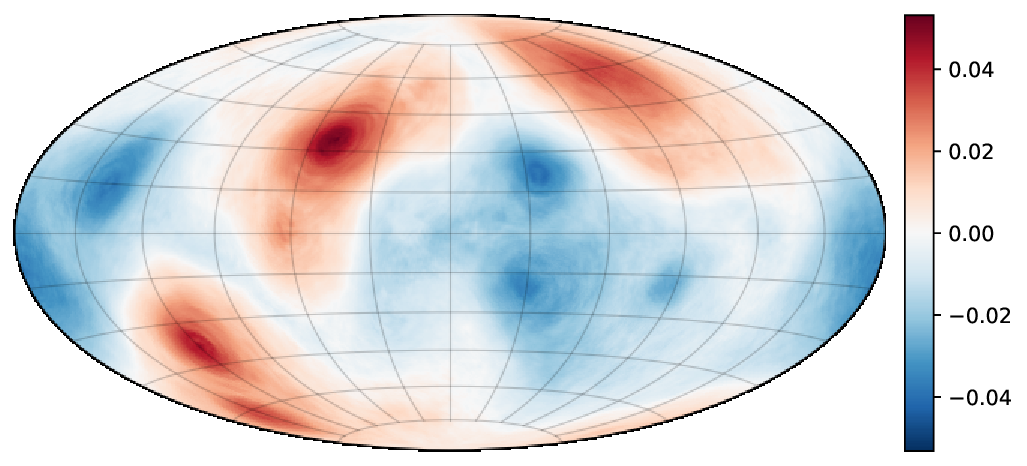} 
    \end{minipage}
    \begin{minipage}[t]{0.5\textwidth}
    % \captionsetup{width=1\textwidth,labelformat=empty} %
      \centering {$\psi$ field,  $t=85$ \phantom{hejhejhej1}}
      \includegraphics[width=\linewidth]{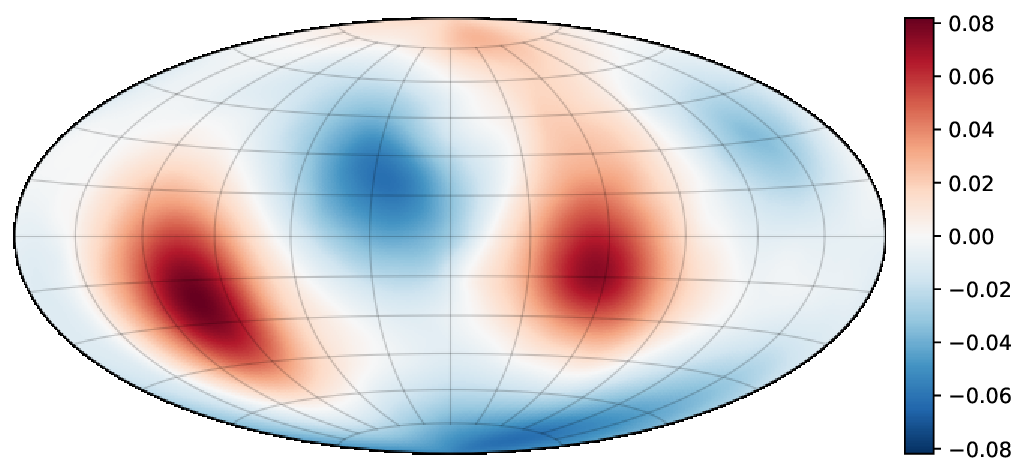} 
    \end{minipage}%
    \hfill % maximize separation between the minipages
    \begin{minipage}[t]{0.5\textwidth}
    % \captionsetup{width=1\textwidth,labelformat=empty}
      \centering {$\theta$ field,  $t=85$ \phantom{hejhejhej1}}
      \includegraphics[width=\linewidth]{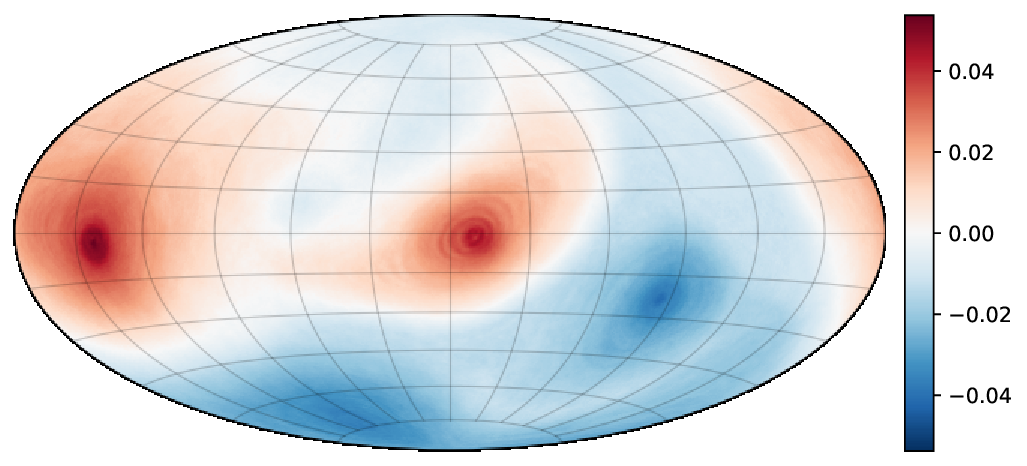} 
    \end{minipage}%
    \hfill % maximize separation between the minipages
    \begin{minipage}[t]{0.5\textwidth}
    % \captionsetup{width=1\textwidth,labelformat=empty}
      \centering {$\psi$ field,  $t=562$ \phantom{hejhejhej1}}
      \includegraphics[width=\linewidth]{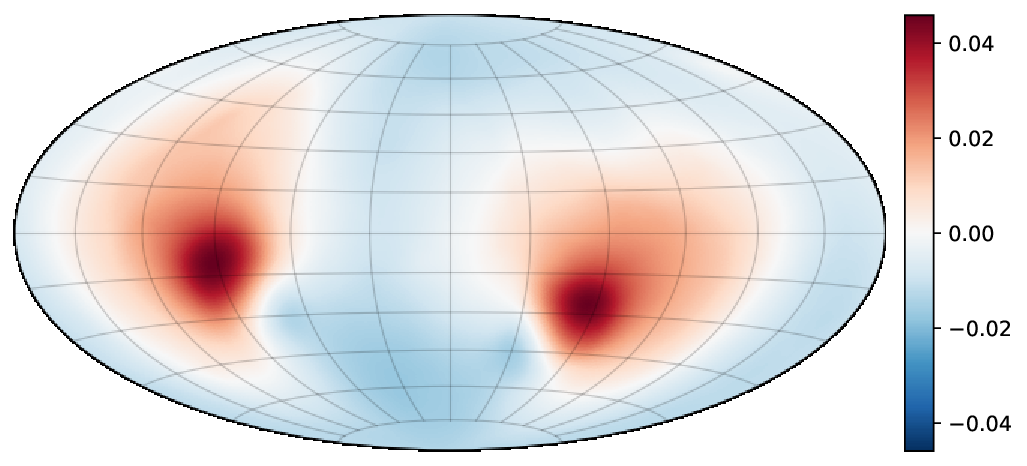} 
    \end{minipage}%
    \hfill % maximize separation between the minipages
    \begin{minipage}[t]{0.5\textwidth}
    % \captionsetup{width=1\textwidth,labelformat=empty}
      \centering {$\theta$ field,  $t=562$ \phantom{hejhejhej1}}
      \includegraphics[width=\linewidth]{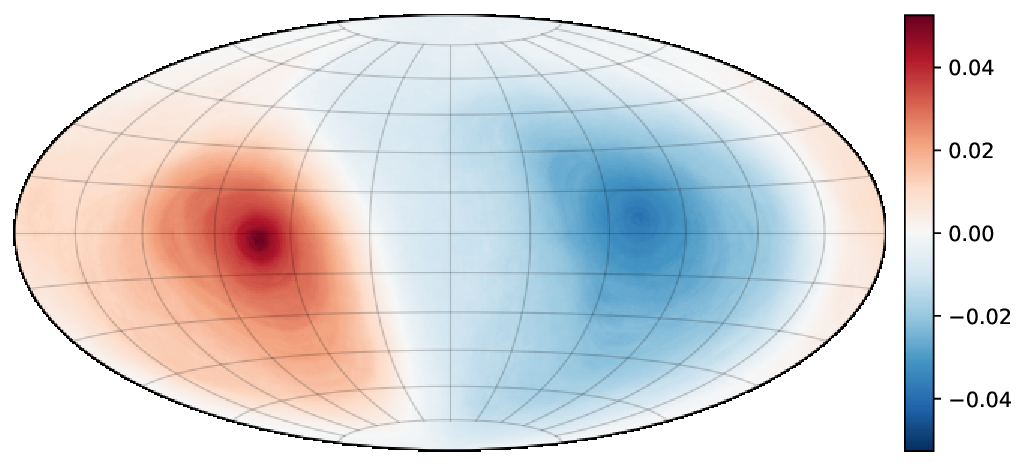} 
    \end{minipage}
\caption{\textbf{Hazeltine:} Evolution of the velocity stream function $\psi(t)$ (left) and the magnetic potential $\theta(t)$ (right). 
The magnetic potential $\theta$ develops into the dipole configuration through intermediate mixing. 
The stream function $\psi$ develops large-scale structures resembling those in the vorticity $\omega$ dynamics.}
\label{magn_pot_dyn_alf}
\end{figure}

Analogously to the simulation in Figure~\ref{gen-magn-evol-figure} for the RMHD model, we present the corresponding evolution of the fields $\theta(t)$ and $\psi(t)$ for the Hazeltine model in Figure~\ref{magn_pot_dyn_alf}. The qualitative dynamics of the magnetic potential $\theta(t)$ in the Hazeltine model is similar to that in RMHD, although it is transported by a different field. One can see from Figure~\ref{magn_pot_dyn_alf} formation of the magnetic dipole through intermediate mixing. 

\begin{figure}
\centering
\includegraphics{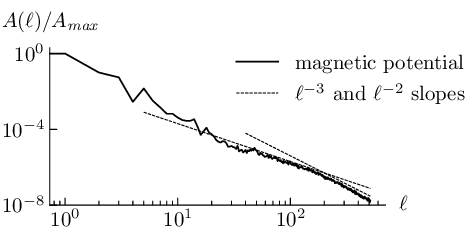}
\caption{\textbf{Hazeltine:} Mean square magnetic potential spectrum of the final state at $t=562$. The observed inverse cascade is consistent with the build-up of the magnetic dipole seen in Figure~\ref{magn_pot_dyn_alf}.}
\label{magn_pot_spectrum_alf}
\end{figure}

The spectral properties of the mean square magnetic potential are different from those in RMHD model. Figure~\ref{magn_pot_spectrum_alf} shows that the scaling for $A$ is $l^{-3}$ with an interval of frequencies $20\lesssim l \lesssim100$ where the scaling is $l^{-2}$. This still indicates the inverse cascade of the mean square magnetic potential, which reflects the formation of large scale magnetic eddies shown in Figure~\ref{magn_pot_dyn_alf}.

\begin{figure}
    \begin{minipage}[t]{0.5\textwidth}
    % \captionsetup{width=1\textwidth,labelformat=empty} %
    \centering {$\chi$ field,  $t=0$ \phantom{hejh}}
      \includegraphics[width=\linewidth]{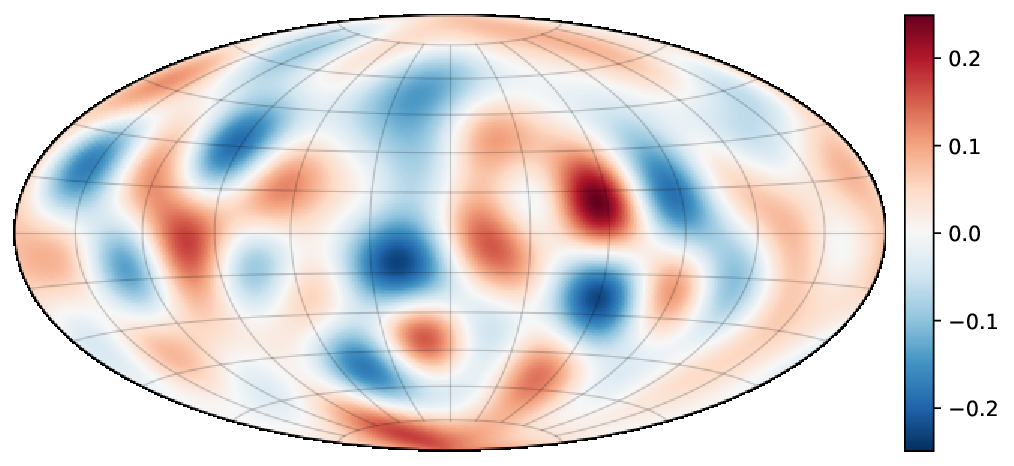} 
      
    \end{minipage}%
    \hfill % maximize separation between the minipages
    \begin{minipage}[t]{0.5\textwidth}
    % \captionsetup{width=1\textwidth,labelformat=empty}
    \centering {$\chi$ field,  $t=3$ \phantom{hejh}}
      \includegraphics[width=\linewidth]{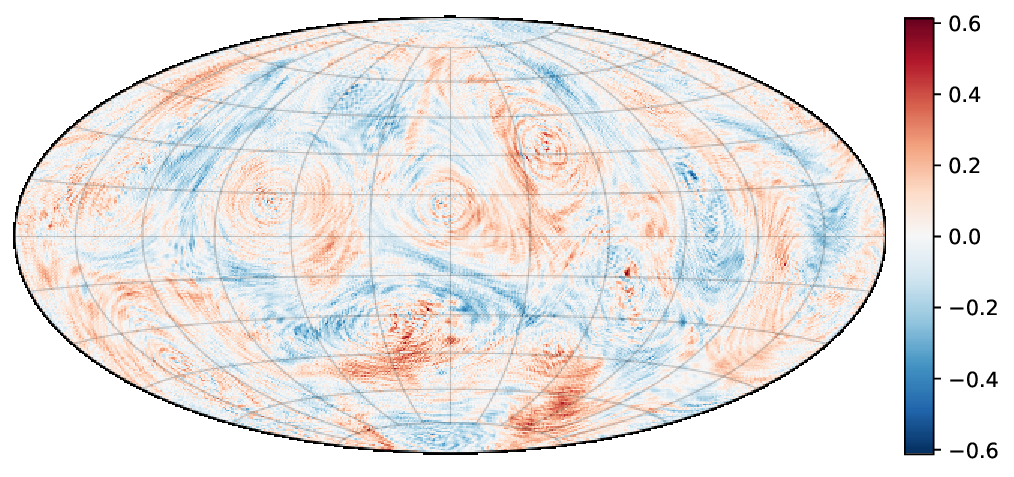} 
      
    \end{minipage}%
    \hfill % maximize separation between the minipages
    \begin{minipage}[t]{0.5\textwidth}
    % \captionsetup{width=1\textwidth,labelformat=empty}
    \centering {$\chi$ field,  $t=203$ \phantom{hejh}}
      \includegraphics[width=\linewidth]{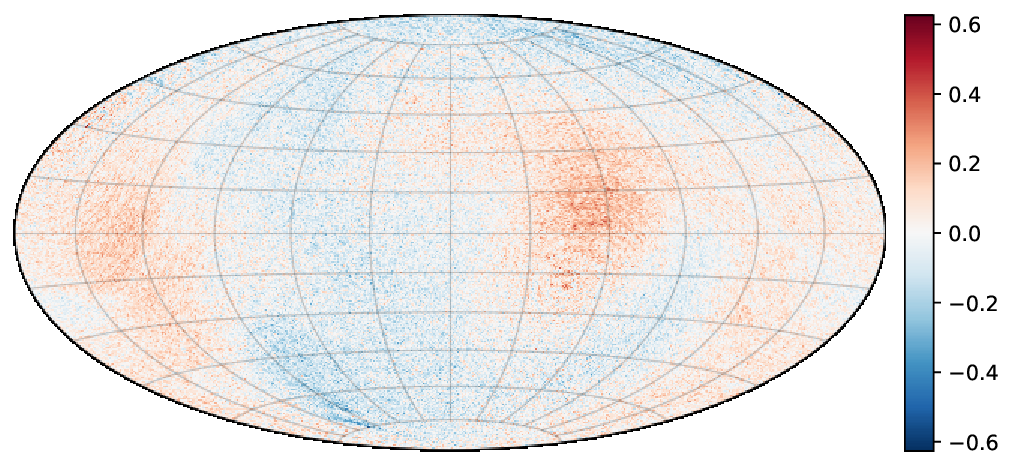} 
      
    \end{minipage}%
    \hfill % maximize separation between the minipages
    \begin{minipage}[t]{0.5\textwidth}
    % \captionsetup{width=1\textwidth,labelformat=empty}
    \centering {$\chi$ field,  $t=562$ \phantom{hejh}}
      \includegraphics[width=\linewidth]{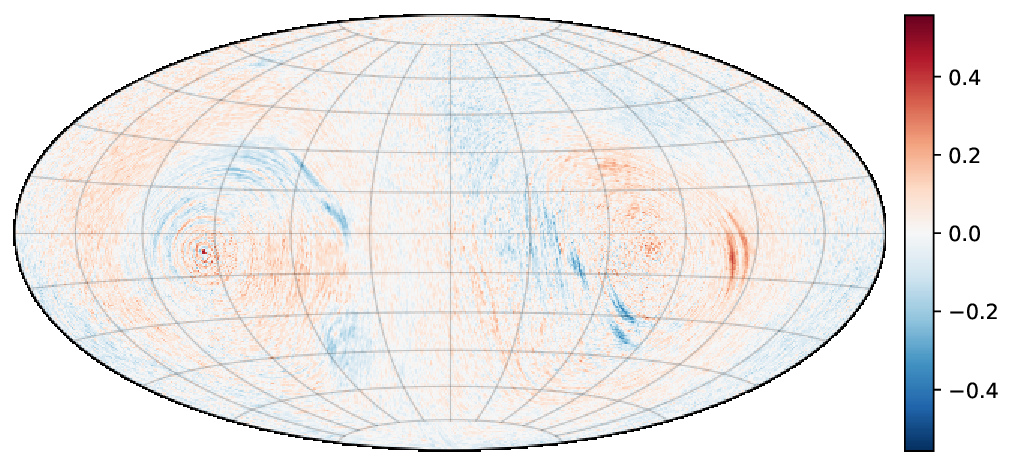} 
      
    \end{minipage}
\caption{\textbf{Hazeltine:} Evolution of the density variation field $\chi(t)$. There are no clear large scale structures remaining at large times, although we observe some circulating filament-like structures centred about the locations of magnetic blobs in Figure~\ref{magn_pot_dyn_alf}.}
\label{chi_field_dyn_alf}
\end{figure}

We conclude this section with the dynamics of the density variation field $\chi(t)$. Although the $\chi(t)$ field fulfils the same equation as the field $\omega(t)$ in the RMHD model, one does not observe a vast amplification of $\chi(t)$ in the Hazeltine model, as seen from Figure~\ref{chi_field_dyn_alf}. In the long-time dynamics, one observes circulations centred approximately at locations of the magnetic potential eddies.

\section{Conclusion and outlook}
\label{sect_concl}
In this paper, we have investigated the long-time behaviour of magnetized ideal two-dimensional fluids governed by Hazeltine's equations and by their two limiting subsystems, the reduced MHD equations and the CHM model. We highlighted the Hamiltonian structure of the three models and presented a structure preserving discretization for numerical simulations based on Zeitlin's matrix equations and a geometric time integrator compatible with the semidirect product Lie algebras underlying the matrix equations. The long-time simulations indicate distinct qualitative features in the long-time behaviour of the mentioned models with the most intricate being the presence of the inverse kinetic energy cascade in the inclusive Hazeltine model, illustrated by the formation of large scale vorticity blobs and by the energy spectra. This behaviour contrasts with the RMHD dynamics where one observes the direct energy cascade accompanied by the vorticity amplification and small scales development. In both RMHD and Hazeltine's models, one observes the formation of large scale magnetic eddies, a magnetic dipole, which is in agreement with the presented mean square of the magnetic potential spectrum. 

We thus conclude that reduced MHD models possess development of large scale structures in some but not all fields.
Thereby, the situation is both analogous to and different from the incompressible 2-D Euler equations, for which emergence of large scale structures and the inverse energy cascade are widely discussed in the literature, cf.~\citep{MaBe2002,BoEc2012}.

Future research can be directed to a deeper theoretical understanding of the inverse cascades in the Hazeltine model, as well as analysis of the importance of the presented geometric discretization for the MHD turbulence studies and its comparison to non-geometric discretizations. Another promising direction could be investigation of MHD equilibria for models with dissipation and adaptation of geometric methods to such models.
\bibliographystyle{abbrvnat}
\bibliography{main.bib}

\end{document}